\documentclass[aps]{revtex4}
\usepackage{amsfonts}
\usepackage{amsmath}
\usepackage{amssymb,epsf}

\begin{document}

\title{Modified TOV in gravity's rainbow:\\
properties of neutron stars and dynamical stability conditions}
\author{S. H. Hendi$^{1,2}$\footnote{%
email address: hendi@shirazu.ac.ir}, G. H. Bordbar$^{1,3}$\footnote{%
email address: ghbordbar@shirazu.ac.ir}, B. Eslam Panah$^{1}$\footnote{%
email address: behzad.eslampanah@gmail.com} and S. Panahiyan$^{1,4}$\footnote{%
email address: sh.panahiyan@gmail.com}} \affiliation{$^1$ Physics
Department and Biruni Observatory, College of Sciences, Shiraz
University, Shiraz 71454, Iran\\
$^2$ Research Institute for Astronomy and Astrophysics of Maragha (RIAAM),
P.O. Box 55134-441, Maragha, Iran\\
$^3$ Center for Excellence in Astronomy and Astrophysics
(CEAA-RIAAM)-Maragha, P.O. Box 55134-441, Maragha 55177-36698, Iran\\
$^4$ Physics Department, Shahid Beheshti University, Tehran 19839, Iran}

\begin{abstract}
In this paper, we consider a spherical symmetric metric to extract the
hydrostatic equilibrium equation of stars in $(3+1)-$dimensional gravity's
rainbow in the presence of cosmological constant. Then, we generalize the
hydrostatic equilibrium equation to $d$-dimensions and obtain the
hydrostatic equilibrium equation for this gravity. Also, we obtain the
maximum mass of neutron star using the modern equations of state of neutron
star matter derived from the microscopic calculations. It is notable that,
in this paper, we consider the effects of rainbow functions on the diagrams
related to the mass-central mass density ($M$-$\rho _{c}$) relation and also
the mass-radius ($M$-$R$) relation of neutron star. We also study the
effects of rainbow functions on the other properties of neutron star such as
the Schwarzschild radius, average density, strength of gravity and
gravitational redshift. Then, we apply the cosmological constant to this
theory to obtain the diagrams of $M$-$\rho _{c}$ (or $M$-$R$) and other
properties of these stars. Next, we investigate the dynamical stability
condition for these stars in gravity's rainbow and show that these stars
have dynamical stability. We also obtain a relation between mass of neutron
stars and Planck mass. In addition, we compare obtained results of this
theory with the observational data.
\end{abstract}

\maketitle

\section{Introduction}

In generalization from Galilean Relativity to Special Relativity, one has to
modify the kinematic equations of Galilean Relativity to obtain an invariant
velocity scale. The same method could be followed in order to derive a
theory, containing an invariant energy scale which is Planck scale. Such
modification is known as the doubly special relativity \cite{MagueijoS2004}.
In other words, we take two upper limits for the particles probing a
spacetime into account: the speed of light and the Planck energy. Due to
modifications in the kinematic structure of the doubly special relativity,
the energy-momentum conservation laws and the Lorentz symmetry group are
altered. There are several theories suggesting that standard energy-momentum
dispersion relation is modified in limit of the Planck scale which among
them one can name; string theory \cite{KosteleckyS1989}, loop quantum
gravity \cite{GambiniP1999} and non-commutative geometry \cite{CarrollHKLO}.
The generalization of the doubly special relativity to incorporate curvature
is known as gravity's rainbow \cite{MagueijoS2004}.

The consideration of doubly special relativity, hence gravity's rainbow, is
related to the effects of quantum spacetime. In other words, it is proposed
that the classical theory of the gravity is an emerging one from quantum
degrees of freedom which in result causes the theory to be an effective one
\cite{BarceloVL,BarceloVL2005,Oriti2007,Gielen2013,Gielen2014}. Such an
effective theory has an energy dependent metric in order to describe
different phenomena \cite{LafranceM1995}. The construction of the metric is
determined by a particle probing it \cite{PengW2008}.

On the other hand, the Einstein theory of gravity has fundamental problem in
UV limit which indicates that it requires modification. It is not necessary
to modify the action in order to have UV completion theory. Alteration of
the metric describing the spacetime in some cases would be sufficient. Such
an approach is employed in Horava-Lifshitz theory of the gravity as well.
The gravity's rainbow also enjoys this feature. In other words, it is an UV
completion theory which in IR limit reduces to classical theory of the
Einstein gravity. It is worthwhile to mention that by considering proper
values for the energy functions of the gravity's rainbow, this theory would
yield a model of the Horava-Lifshitz which indicates that these theories are
related. The UV completion theories require that their energy-momentum
dispersion relation to be modified. Such modification is also observed in
discrete spacetime \cite{Hooft1996}, spacetime foam \cite{AmelinoEMNS},
spin-network in loop quantum gravity (LQG) \cite{GambiniP1999}, ghost
condensation \cite{Faizal2011} and non-commutative geometry \cite%
{CarrollHKLO}. In addition, experimental observations also confirmed that
such modification must be considered in the UV limit \cite{Abraham2010}.

Several studies are conducted in context of black holes by consideration of
the gravity's rainbow \cite%
{GarattiniM2014,Ali2014,GimK2015,HendiFEP2015,Hendi2015,HendiPEM2016}. It
was shown that the thermodynamic properties of black holes are modified in
gravity's rainbow in such a way that it admits the existence of remnant for
black holes. In evaporation of the black holes in gravity's rainbow, the
temperature of black holes goes to zero while its size is finite. Therefore,
the black hole is not fully evaporated. Due to this feature, it is proposed
that this model could provide a solution regarding the information paradox
\cite{AliFM2015} and the formation of naked singularity at the last stage of
the black hole's evaporation. The existence of remnant is observed for Kerr
black holes, Kerr-Newman black holes in de Sitter space, charged AdS black
holes, higher dimensional Kerr-AdS black holes and black saturn \cite%
{AliFK2015}. In addition, due to this phenomena, there exists an energy
limit in which the mini black holes can be produced at LHC \cite{AliFK2015b}%
. It is worthwhile to mention that the gravity's rainbow also holds the
usual uncertainty principle \cite{LingLZ2007,LiLH2009}. Recently, there has
been an increasing interest in gravity's rainbow \cite%
{GarattiniMa2014,HendiF2015,ChangW2015,SantosGA2015,AliK2015}. The wormhole
solutions in the context of gravity's rainbow have been investigated in Ref.
\cite{GarattiniL201512}. A study regarding the extended informal approach in
gravity's rainbow is done in Ref. \cite{CarvalhoLB2015}. In addition, the
rainbow metric from quantum has been obtained in Ref. \cite%
{AssanioussiDL2015}. On the other hand, the effects of gravity's rainbow on
the thermodynamics of different models of black holes have been investigated
in literature \cite%
{GarattiniM2014,Ali2014,GimK2015,HendiFEP2015,Hendi2015,HendiPEM2016}.
Einstein theory of gravity is an effective theory which confronts specific
problems in different regimes such as UV or in describing some phenomena
such as increasing accelerating expansion of the Universe which was proven
by the observation of high red-shift supernova \cite%
{Perlmutter1999,Perlmutter1999b,Riess2014} and the measurement of angular
fluctuations of cosmic microwave background fluctuations \cite%
{Lee2001,Netterfield2002,Halverson2002,Spergel2003}. One of the best
candidates for modifications of Einstein theory is adding the cosmological
constant to Einstein's Lagrangian \cite{Padmanabhan2003,FriemanTH2008}.

The physics governing the structure of stars includes the interaction
between gravitational force and internal pressure which results into an
equilibrium state. This is known as hydrostatic equilibrium equation (HEE)
which plays the essential role in studying the structure of stars. On the
other hand, it was proven that in studying the compact objects such as the
neutron and strange quark stars, consideration of curvature of space-time
and generalization of Newtonian gravity to general relativity is necessary
for describing different phenomena and making accurate predictions. The
first attempt for obtaining Einsteinian HEE for stars was done by Tolman,
Oppenheimer and Volkoff (TOV) \cite{Tolman1934,Tolman1939,OppenheimerV1939}.
The TOV approach toward studying the structure of the stars has been
employed by several authors \cite%
{BordbarM1998,YunesV2003,SilbarR2004,NarainSM2006,BordbarBY2006,BoonsermVW2007,LiWC2012,OliveiraVFS2014,HeFLN2015}%
. In addition, the generalization to modified theories of gravity such as $%
F(R)$, $F(G)$ \cite{AstashenokCOa,AstashenokCOb,MomeniGMM2015,AbbasMAMQ2015}
and dilaton \cite{HendiBEN2015} gravities are done (for more details see
Refs. \cite%
{MeyerPB2012,AstashenokCO2013,OrellanaGPR2013,ArbanilLZ2013,GoswamiNMG2014,LemosLQZ2015}%
).

There has been an ongoing debate regarding the consideration of the higher
dimensionality in studying different phenomena. There are several reasons
which are supporting such generalization: First of all, in context of
particles interactions, the necessity of such generalization is expressed
for having consistent theories. On the other hand, it is essential to
consider higher dimensionality in order to have an effective theory of
superstring \cite{GreenS1984,GreenS1985,CandelasHSW1985}. The phenomenology
of higher dimensionality has proven to be richer \cite%
{HoravaW1996,LukasOW1999,RandallS1999}. As for the stars, a study regarding
higher dimensional mass-radius relation for a star with the uniform density
was done in Ref. \cite{Paul2001}. The consideration of Kaluza-Klein model
for describing the neutron stars was done in Ref. \cite{LiddleMH}. This
study highlighted the importance of higher dimensionality. In addition, the
relativistic anisotropic stars were investigated, and the effects of higher
dimensions were pointed out \cite{PaulCK2014}. The TOV equation of Einstein-$%
\Lambda $ gravity in higher dimensions was studied in the Refs. \cite%
{PonceC2000,BordbarHE2015}. In this paper, we are interested in obtaining
HEE for Einstein-$\Lambda $ in gravity's rainbow and its generalization to $d
$-dimensional case.

The outline of the paper will be as follows. First, we study a spherical
symmetric metric and extract the HEE in Einstein-$\Lambda $ gravity's
rainbow for $(3+1)-$dimensions. Next, we obtain a global equation of
hydrostatic equilibrium for compact stars in the higher dimensions in this
gravity. The maximum mass for neutron star in the presence and absence of
cosmological constant in gravity's rainbow is investigated. In the next
section, we compare results obtained for theory with observational compact
object. The last section is devoted to closing remarks.

\section{$(3+1)-$dimensional HEE in gravity's rainbow}

\label{3+1dimRainbow}

Here, we present the Einstein gravity with the cosmological constant. The
action of this gravity is given by
\begin{equation}
I_{G}=-\frac{1}{16\pi }\int_{\mathcal{M}}d^{d}x\sqrt{-g}\{R-2\Lambda
\}+I_{Matt},  \label{actionEN}
\end{equation}%
where $R$ is the Ricci scalar, $\Lambda $ is the cosmological constant and $%
I_{Matt}$ is the action of matter field. Varying the action (\ref{actionEN})
with respect to the metric tensor $g_{\mu }^{\nu }$, the equation of motion
for this gravity can be written as%
\begin{equation}
G_{\mu }^{\nu }+\Lambda g_{\mu }^{\nu }=KT_{\mu }^{\nu },  \label{EqEN}
\end{equation}%
where $K=\frac{8\pi G}{c^{4}}$ and also, $G_{\mu }^{\nu }$ and $T_{\mu
}^{\nu }$ are the Einstein and energy-momentum tensors, respectively.

Here, we intend to obtain the static solutions of Eq. (\ref{EqEN}) in
gravity's rainbow. To do so, we use the metric which is energy dependent. As
it was pointed out, one of the basics for gravity's rainbow is deformation
of the standard energy-momentum relation%
\begin{equation}
E^{2}L^{2}(\varepsilon )-p^{2}H^{2}(\varepsilon )=m^{2},
\end{equation}%
in which $\varepsilon =\frac{E}{E_{P}}$ where $E_{P}$ is the Planck energy.
Considering the mentioned upper limit for energies that a particle can
obtain, we have%
\begin{equation}
\varepsilon \leq 1.
\end{equation}

The functions $L(\varepsilon )$ and $H(\varepsilon )$ are rainbow functions
which satisfy the following conditions,%
\begin{equation}
\lim_{\varepsilon \rightarrow 0}L(\varepsilon )=1,~\ \ \ \ \ \ \ \
\lim_{\varepsilon \rightarrow 0}H(\varepsilon )=1.
\end{equation}

These conditions ensure the standard energy-momentum relation in the
infrared limit. Now, we are in a position to construct the energy dependent
metric as follows \cite{PengW2008}%
\begin{equation}
h(\varepsilon )=\eta ^{\mu \nu }e_{\mu }(\varepsilon )\otimes e_{\nu
}(\varepsilon ),
\end{equation}%
where%
\begin{equation}
e_{0}(\varepsilon )=\frac{1}{L(\varepsilon )}e_{0}^{\sim },~\ \ \ \ \ \ \ \
e_{i}(\varepsilon )=\frac{1}{H(\varepsilon )}e_{i}^{\sim },
\end{equation}%
in which the tilde quantities refer to the energy independent frame fields.
Therefore, one can assume a spherical symmetric space-time in the following
form,%
\begin{equation}
ds^{2}=\frac{f(r)}{L^{2}(\varepsilon )}dt^{2}-\frac{1}{H^{2}(\varepsilon )}%
\left( \frac{dr^{2}}{g(r)}+r^{2}\left( d\theta ^{2}+\sin ^{2}\theta d\varphi
^{2}\right) \right) ,  \label{Metric}
\end{equation}%
where $f(r)$ and $g(r)$ are radial dependent functions which should be
determined. In addition, $L(\varepsilon )$ and $H(\varepsilon )$ are the
energy functions which will be introduced later.

The energy-momentum tensor for a perfect fluid is%
\begin{equation}
T^{\mu \nu }=\left( c^{2}\rho +P\right) U^{\mu }U^{\nu }-Pg^{\mu \nu },
\label{EMTensorEN}
\end{equation}%
where $\rho $ and $P$ are density and pressure of the fluid which are
measured by local observer, respectively, and $U^{\mu }$ is defined as
\begin{equation}
U^{\mu }=\left( \frac{L(\varepsilon )}{\sqrt{f(r)}},0,0,0\right) ,
\end{equation}%
which is the fluid four-velocity with following restriction
\begin{equation}
g_{\mu \nu }U^{\mu }U^{\nu }=1.
\end{equation}

Using Eq. (\ref{EMTensorEN}) and the metric introduced in Eq. (\ref{Metric}%
), we can obtain the components of energy-momentum tensor for $(3+1)$%
-dimensions as follows%
\begin{equation}
T_{0}^{0}=\rho c^{2}~\ \ \ \ \&~\ \ \ \ T_{1}^{1}=T_{2}^{2}=T_{3}^{3}=-P.
\label{4dim}
\end{equation}

We consider the metric (\ref{Metric}) and Eq. (\ref{4dim}) for perfect fluid
and obtain the components of Eq. (\ref{EqEN}) with the following forms%
\begin{eqnarray}
Kc^{2}r^{2}\rho &=&\Lambda r^{2}+\left( 1-g\right) H^{2}(\varepsilon )-rg{%
^{\prime }}H^{2}(\varepsilon ),  \label{1} \\
Kr^{2}fP &=&-\Lambda r^{2}f-\left( 1-g\right) H^{2}(\varepsilon
)f+rgH^{2}(\varepsilon )f{^{\prime },}  \label{2} \\
4Krf^{2}P &=&-4\Lambda rf^{2}+2\left( gf\right) {^{\prime }H^{2}(\varepsilon
)f+r}\left[ g{^{\prime }}f{^{\prime }+2g}f{^{\prime \prime }}\right]
H^{2}(\varepsilon ){f-rgH^{2}(\varepsilon )f^{\prime 2}},  \label{3}
\end{eqnarray}%
where $f$, $g$, $\rho $ and $P$ are functions of $r$. We note that the prime
and double prime denote the first and second derivatives with respect to $r$%
, respectively.

Using Eqs. (\ref{1}-\ref{3}) and after some calculations, we have
\begin{equation}
\frac{dP}{dr}+\frac{\left( c^{2}\rho +P\right) f{^{\prime }}}{2f}=0.
\label{extraEQ}
\end{equation}

Now, we obtain $f{^{\prime }}$ from Eq. (\ref{2}) as follow
\begin{equation}
f{^{\prime }}=\frac{\left[ r^{2}\left( \Lambda +KP\right) +\left( 1-g\right)
H^{2}(\varepsilon )\right] f}{rgH^{2}(\varepsilon )}.  \label{diff(r)}
\end{equation}

Then, for obtaining $g$, we use Eq. (\ref{1})
\begin{equation}
g=1+\frac{\Lambda }{3H^{2}(\varepsilon )}r^{2}-\frac{c^{2}KM_{eff}}{4\pi r},
\label{4g(r)}
\end{equation}%
where $M_{eff}$ is a function of $r$ and $\varepsilon $: $M_{eff}\left(
r,\varepsilon \right) =\int \frac{4\pi r^{2}\rho (r)}{H^{2}(\varepsilon )}dr$%
. It is notable that the obtained relation for effective mass is a function
of radial coordinate and energy functions. This is a direct contribution of
the gravity's rainbow. By inserting Eqs. (\ref{diff(r)}) and (\ref{4g(r)})
in Eq. (\ref{extraEQ}), we can extract the HEE in Einstein-$\Lambda $
gravity's rainbow for $(3+1)$-dimension as%
\begin{equation}
\frac{dP}{dr}=\frac{\left[ 3c^{2}GM_{eff}H^{2}(\varepsilon )+r^{3}\left(
\Lambda c^{4}+12\pi GP\right) \right] \left( c^{2}\rho +P\right) }{c^{2}r%
\left[ 6GM_{eff}H^{2}(\varepsilon )-c^{2}r\left( \Lambda
r^{2}+3H^{2}(\varepsilon )\right) \right] }.  \label{TOV}
\end{equation}

We see that this equation is different from HEE, and for $\Lambda =0$ and $%
H(\varepsilon )=1$, it yields the usual TOV equation \cite%
{Tolman1934,Tolman1939,OppenheimerV1939}. Also, for $\Lambda \neq 0$ and $%
H(\varepsilon )=1$, it reduces to HEE obtained in Ref. \cite{BordbarHE2015}.

Considering different phenomenologies, one can employ three different cases
for rainbow functions.

Motivated by studies conducted in loop quantum gravity and non-commutative
geometry, the first case has the following energy functions, \cite%
{JacobMAP2010,Amelino-Camelia2013}%
\begin{equation}
L(\varepsilon )=1\ \ \ \ \ \ \&\ \ \ \ \ H(\varepsilon )=\sqrt{1-\eta
\varepsilon ^{n}}.  \label{caseI}
\end{equation}%
Using Eqs. (\ref{TOV}) and (\ref{caseI}), the HEE is%
\begin{equation}
\frac{dP}{dr}=\frac{\left[ 3c^{2}GM_{eff}\left[ 1-\eta \varepsilon ^{n}%
\right] +r^{3}\left( \Lambda c^{4}+12\pi GP\right) \right] \left( c^{2}\rho
+P\right) }{c^{2}r\left[ 6GM_{eff}\left[ 1-\eta \varepsilon ^{n}\right]
-c^{2}r\left( \Lambda r^{2}+3\left[ 1-\eta \varepsilon ^{n}\right] \right) %
\right] },  \label{TOVI}
\end{equation}%
where for $\eta =0$, Eq. (\ref{TOVI}) reduces to the HEE obtained for
Einstein-$\Lambda $ gravity in Ref. \cite{BordbarHE2015}. \textbf{Here,
since }$\ H(\varepsilon )$\textbf{\ is a square root function, one can find
following condition regarding its parameters}%
\begin{equation*}
\eta <\frac{1}{\varepsilon ^{n}}.
\end{equation*}

The second case is related to the hard spectra from gamma-ray bursts which
has energy functions as \cite{AmelinoEMNS}%
\begin{equation}
L(\varepsilon )=\frac{e^{\beta \varepsilon }-1}{\beta \varepsilon }\ \ \ \ \
\ \&\ \ \ \ \ H(\varepsilon )=1.  \label{caseII}
\end{equation}

It is notable that due to the structure of obtained TOV equation (Eq. (\ref%
{TOV})), this case of energy functions will lead to absence of the effects
of gravity's rainbow. In other words, for this case, the TOV equation is
independent of energy functions which is not of our interest.

The third case of energy functions is due to consideration of constancy of
the velocity of light which leads to the following relation, \cite%
{MagueijoS2002}%
\begin{equation}
L(\varepsilon )=H(\varepsilon )=\frac{1}{1-\lambda \varepsilon }.
\label{caseIII}
\end{equation}

We use Eq. (\ref{caseIII}) for obtaining the HEE, so we have%
\begin{equation}
\frac{dP}{dr}=\frac{\left[ 3c^{2}GM_{eff}\left( \frac{1}{1-\lambda
\varepsilon }\right) ^{2}+r^{3}\left( \Lambda c^{4}+12\pi GP\right) \right]
\left( c^{2}\rho +P\right) }{c^{2}r\left[ 6GM_{eff}\left( \frac{1}{1-\lambda
\varepsilon }\right) ^{2}-c^{2}r\left( \Lambda r^{2}+3\left( \frac{1}{%
1-\lambda \varepsilon }\right) ^{2}\right) \right] },  \label{TOVII}
\end{equation}%
where for $\lambda =0$, Eq. (\ref{TOVII}) reduces to
Einstein-$\Lambda $ gravity \cite{BordbarHE2015}. Since the energy
functions must be positive valued (for avoiding the change of
metric signature), one can extract following condition for this
model of gravity's rainbow

\begin{equation*}
\lambda <\frac{1}{\varepsilon }.
\end{equation*}

\section{HEE of gravity's rainbow in higher dimensions}

\label{higherdimensions}

Now, we are going to obtain the HEE in Einstein-$\Lambda $ gravity's rainbow
for higher dimensions ($d\geq 5$ where $d$ represents the dimensionally of
space-time). We consider the following metric%
\begin{equation}
ds^{2}=\frac{f(r)}{L^{2}(\varepsilon )}dt^{2}-\frac{1}{H^{2}(\varepsilon )}%
\left( \frac{dr^{2}}{g(r)}+r^{2}d\Omega _{k}^{2}\right) ,  \label{Metricd}
\end{equation}%
where%
\begin{equation}
d\Omega _{k}^{2}=d\theta
_{1}^{2}+\sum\limits_{i=2}^{d-2}\prod\limits_{j=1}^{i-1}\sin ^{2}\theta
_{j}d\theta _{i}^{2}.
\end{equation}

Then, we must obtain Eq. (\ref{4dim}) for arbitrarily dimensions, therefore
we have,%
\begin{equation}
T_{0}^{0}=c^{2}\rho ~\ \ \ \ \&~\ \ \ \
T_{1}^{1}=T_{2}^{2}=T_{3}^{3}=...=T_{d-1}^{d-1}=-P.  \label{ddim}
\end{equation}

We use the above equation to obtain a global relation for the HEE in higher
dimensions for Einstein-$\Lambda $ gravity's rainbow. Using Eqs. (\ref{EqEN}%
) and (\ref{ddim}) for the metric (\ref{Metricd}), the components of the
equation of motion for Einstein-$\Lambda $ gravity's rainbow can be written
as%
\begin{eqnarray}
K_{d}c^{2}r^{2}\rho  &=&\Lambda r^{2}+\frac{(d-2)(d-3)}{2}\left( 1-g\right)
H^{2}(\varepsilon )-\frac{(d-2)}{2}rg{^{\prime }}H^{2}(\varepsilon ),
\label{d1} \\
K_{d}r^{2}fP &=&-\Lambda r^{2}f-\frac{(d-2)(d-3)}{2}\left( 1-g\right)
H^{2}(\varepsilon )f+\frac{(d-2)}{2}rgH^{2}(\varepsilon ){f{^{\prime }},}
\label{d2} \\
4K_{d}rf^{2}P &=&-4\Lambda rf^{2}-\frac{2\left( d-3\right) \left( d-4\right)
}{r}\left( 1-g\right) H^{2}(\varepsilon )f^{2}+r\left[ g{^{\prime }}f{%
^{\prime }+2g}f{^{\prime \prime }}\right] H^{2}(\varepsilon )f  \notag \\
&&+2\left( d-3\right) \left( gf\right) {^{\prime }}H^{2}(\varepsilon
)f-rgH^{2}(\varepsilon )f{^{\prime 2},}  \label{d3}
\end{eqnarray}%
in which $K_{d}=\frac{8\pi G_{d}}{c^{4}}$ and $G_{d}$ is the gravitational
constant in $d-$dimensons and $G_{d}$ is defined as $G_{d}=GV_{d-4}$, where $%
G$ denotes the four dimensional gravitational constant and also, $V_{d-4}$
is the volume of extra space. We obtain Eq. (\ref{extraEQ}) in the same
manner as it was described for $(3+1)$-dimensions, then using Eq. (\ref{d2}%
), we get $f{^{\prime }}$ as follows%
\begin{equation}
f{^{\prime }}=\frac{2\left[ r^{2}\left( \Lambda +K_{d}P\right) +\frac{%
(d-2)(d-3)}{2}\left( 1-g\right) H^{2}(\varepsilon )\right] f}{%
rg(d-2)H^{2}(\varepsilon )}.  \label{difff(r)}
\end{equation}

We calculate $g$ of Eq. (\ref{d1}) as
\begin{equation}
g=1+\frac{2\Lambda }{(d-1)(d-2)H^{2}(\varepsilon )}r^{2}-\frac{%
c^{2}K_{d}M_{eff}\left( r,\varepsilon \right) \Gamma \left( \frac{d-1}{2}%
\right) }{\left( d-2\right) \pi ^{\left( d-1\right) /2}r^{d-3}}.
\label{dg(r)}
\end{equation}

It should be noted that we have used $M_{eff}\left( r,\varepsilon \right)
=\int \frac{2\pi ^{(d-1)/2}}{\Gamma ((d-1)/2)H^{2}(\varepsilon )}r^{d-2}\rho
(r)dr$ in the above equation and also $\Gamma $ is the gamma function, which
satisfies some conditions as $\Gamma (1/2)=\sqrt{\pi }$, $\Gamma (1)=1$ and $%
\Gamma (x+1)=x\Gamma (x)$.

Using Eqs. (\ref{difff(r)}) and (\ref{dg(r)}) in Eq. (\ref{extraEQ}), we can
get the HEE in the Einstein-$\Lambda $ gravity for $d$-dimensions%
\begin{equation}
\frac{dP}{dr}=\frac{\left[ \frac{(d-1)(d-3)\Gamma (\frac{d-1}{2}%
)c^{2}K_{d}M_{eff}H^{2}(\varepsilon )}{4\pi ^{\left( d-1\right) /2}r^{d-1}}%
+\left( \Lambda +\frac{\left( d-1\right) K_{d}P}{2}\right) \right] \left(
c^{2}\rho +P\right) }{r\left[ -\Lambda +\frac{(d-1)}{r^{d-1}}\left( \frac{%
\Gamma (\frac{d-1}{2})c^{2}K_{d}M_{eff}}{2\pi ^{^{\left( d-1\right) /2}}}-%
\frac{\left( d-2\right) r^{d-3}}{2}\right) H^{2}(\varepsilon )\right] },
\label{dTOV}
\end{equation}%
where for $(3+1)-$dimensional limit, Eq. (\ref{dTOV}) reduces to Eq. (\ref%
{TOV}).

\section{Structure properties of neutron star}

\subsection{Equation of state of neutron star matter \label{Neutronstar}}

It is stated that one can derive the properties of neutron star matter by
obtaining the equation of state of neutron star matter. The constituents of
the interior part of a neutron star include the neutrons, protons, electrons
and muons which are in charge neutrality and beta equilibrium conditions
(beta-stable matter) \cite{ShapiroT1983}. On the other hand, for studying
the equation of state of stars, there are several approaches which among
them one can name the microscopic constrained variational calculations based
on the cluster expansion of the energy functional \cite%
{BordbarR2002,Bordbar2004}. In this approach, the new Argonne AV$18 $\ and
charged dependent Reid-$93$ are employed as the two-nucleon potentials (see
Refs. \cite{WiringaSS1995,StoksKTd1994}, for more details). Also, a good
convergence, lack of need for any free parameter in formalism and more
accuracy comparing to other semi-empirical parabolic approximation methods
are of the advantages of these methods. The modern nucleon-nucleon
potentials are isospin projection ($Tz$) dependent in which a microscopic
computation of asymmetry energy is carried on for the asymmetric nuclear
matter calculations (see Ref. \cite{BordbarM1998} for more details).

\begin{figure}[tbp]
$%
\begin{array}{c}
\epsfxsize=7cm \epsffile{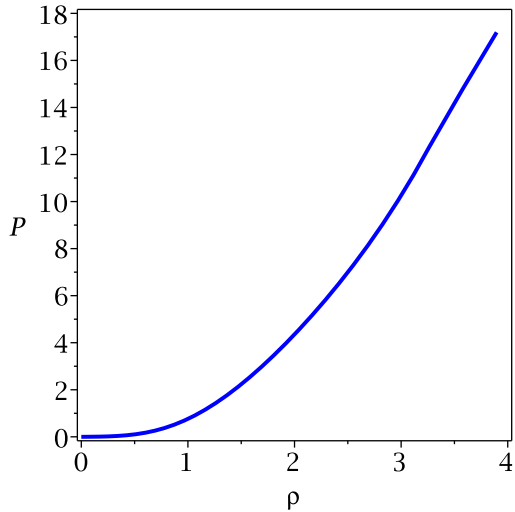}%
\end{array}
$%
\caption{Equation of state of neutron star matter (pressure, $P$ ($10^{35}$
erg/$cm^{3}$) versus density, $\protect\rho $ ($10^{15}$ gr/$cm^{3}$)).}
\label{Fig1}
\end{figure}

In this paper, we employ AV$18$ potential \cite{BordbarR2002,Bordbar2004}
for calculating the modern equation of state for neutron star matter. Then,
we will study some physical properties of neutron star structure. Our result
for obtained equation of state of neutron star matter is presented in Fig. %
\ref{Fig1}. We extract the mathematical forms for the equation of state
presented in Fig. \ref{Fig1} as
\begin{equation}
P=\sum_{i=1}^{7}\mathcal{A}_{i}\rho ^{7-i},  \label{EoS}
\end{equation}%
where $\mathcal{A}_{i}$ are%
\begin{eqnarray*}
\mathcal{A}_{1} &=&-3.518\times 10^{-57}, \\
\mathcal{A}_{2} &=&3.946\times 10^{-41}, \\
\mathcal{A}_{3} &=&-1.67\times 10^{-25}, \\
\mathcal{A}_{4} &=&3.242\times 10^{-10}, \\
\mathcal{A}_{5} &=&-1.458\times 10^{5}, \\
\mathcal{A}_{6} &=&2.911\times 10^{19}, \\
\mathcal{A}_{7} &=&-9.983\times 10^{31}.
\end{eqnarray*}

In order to have a better insight regarding the properties of these neutron
stars, we study both energy and stability conditions in the following
subsections.

\subsubsection{Energy conditions}

Here, we investigate the energy conditions such as the null energy condition
(NEC), weak energy condition (WEC), strong energy condition (SEC) and also,
dominant energy condition (DEC) at the center of neutron star. We have%
\begin{eqnarray}
NEC &\rightarrow &\ P_{c}+\rho _{c}\geq 0,  \label{11} \\
WEC &\rightarrow &\ P_{c}+\rho _{c}\geq 0,\ \ \ \&\ \ \rho _{c}\geq 0,
\label{22} \\
SEC &\rightarrow &\ P_{c}+\rho _{c}\geq 0,\ \ \ \&\ \ 3P_{c}+\rho _{c}\geq 0,
\label{33} \\
DEC &\rightarrow &\ \rho _{c}>\left\vert P_{c}\right\vert ,  \label{44}
\end{eqnarray}%
where $\rho _{c}$ and $P_{c}$ are the density and pressure at the center of
neutron star ($r=0$), respectively. Considering Fig. \ref{Fig1} and the
above conditions (\ref{11}-\ref{44}), our results are given in table \ref%
{tab11}. According to Fig. \ref{Fig1} and table \ref{tab11}, we observe that
all energy conditions are satisfied. So, the equation of state of neutron
star matter introduced in this paper is suitable.

\begin{table}[tbp]
\caption{Energy conditions for neutron star.}
\label{tab11}
\begin{center}
\begin{tabular}{cccccc}
\hline\hline
$\rho_{0} (10^{9}\frac{kg}{cm^3})$ & $P_{0} (10^{9}\frac{kg}{cm^3})$ & $NEC$
& $WEC$ & $SEC$ & $DEC$ \\ \hline\hline
$3895.99$ & $1910.39$ & $\checkmark$ & $\checkmark$ & $\checkmark$ & $%
\checkmark$ \\ \hline
\end{tabular}%
\end{center}
\end{table}

\subsubsection{Stability}

\begin{figure}[tbp]
$%
\begin{array}{c}
\epsfxsize=7cm \epsffile{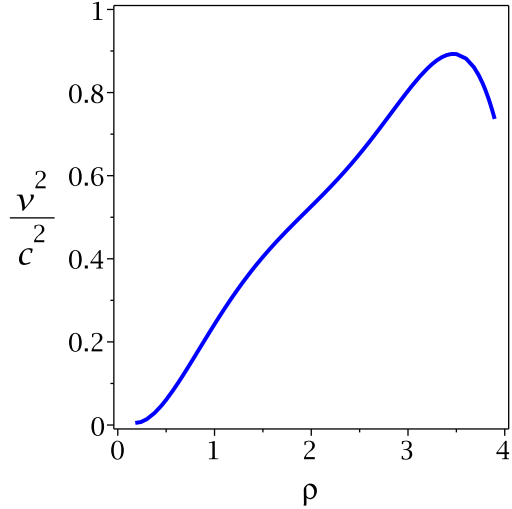}%
\end{array}
$%
\caption{Sound speed ($v^{2}/c^{2}$) versus density ($\protect\rho \times
10^{15}$ (gr/$cm^{3}$)).}
\label{Fig2}
\end{figure}

In order to investigate the stability of equation of state of neutron star
matter for a physically acceptable model, one expects that the velocity of
sound ($v$) be less than the light's velocity ($c$) \cite%
{Herrera1992,AbreuHN2007}. By considering the stability condition ($0\leq
v^{2}=\left( \frac{dP}{d\rho }\right) \leq c^{2}$) and Eq. (\ref{EoS}), and
comparing them with diagrams related to speed of sound-density relationship
plot in Fig. \ref{Fig2}, it is evident that this equation of state of
neutron star matter satisfies the inequality $0\leq v^{2}\leq c^{2}$.

Our investigations show that our equation of state of neutron star matter
satisfies both energy and stability conditions. Now, we focus on
investigation of gravitational mass and radius for neutron stars in
gravity's rainbow.

\subsection{Properties of neutron stars in gravity's rainbow without the
cosmological constant}

Study of compact objects such as neutron stars has been of great interest
for the astrophysicists in the last two decades (see Refs. \cite%
{ThorsettAMT1993,HeylH1998,KovacsCH2010,CowardGSHRLKBB,PejchaTK,RemmenW,KantorGC2016,HoyosFJV}
\ for more details). Of the most important features of neutron stars are the
maximum mass and radius, and determining them. Because, there is a critical
maximum mass in which for masses smaller than that the degeneracy pressure
originated from the nucleons prevents an object from becoming a black hole
\cite{ShapiroT1983}. Therefore, obtaining the maximum gravitational mass of
neutron stars is of great importance in astrophysics. Due to many errors in
direct ways of measuring mass and radius of the neutron stars by
observations of the X-ray pulsars and X-ray bursts, one is not able to
obtain an accurate mass and radius for these stars. On the other hand, using
the binary radio pulsars \cite%
{WeisbergT1984,Liang1986,Heap1992,Quaintrell2003}, leads to highly accurate
results for the mass and radius of neutron stars.

Now, by employing the equation of state of neutron star matter which is
presented in Fig. \ref{Fig1}, and numerical approach for integrating the HEE
obtained in the equation (\ref{TOV}), we obtain the maximum mass and radius
of neutron star. It is worthwhile to mention that the neutron star mass and
radius are central mass density ($\rho _{c}$) dependent in this approach.
For this purpose, one can consider the boundary conditions $P(r=0)=P_{c}$
and $m(r=0)=0$, and integrates Eq. (\ref{TOV}) outwards to a radius $r=R$ in
which $P$ vanishes for selecting a $\rho _{c}$. This leads to the neutron
star radius $R$ and mass $M=m(R)$. The results are presented in various
tables and figures.

The Einstein gravity case has been investigated in Ref. \cite{BordbarH2006},
and the maximum mass of neutron stars by using the modern equations of state
of neutron star matter derived from microscopic calculations was obtained.
It was shown that the maximum mass for neutron stars is about $1.68M_{\odot
} $ in Einstein gravity. The effect of the cosmological constant on maximum
mass of neutron star was investigated, and it was pointed out that
considering the positive values of the cosmological constant, the maximum
mass of neutron star decreases \cite{BordbarHE2015} and also, the behavior
of neutron star (for example; diagram related to the maximum mass-radius)
for the negative values of the cosmological constant are not logical (see
Ref. \cite{BordbarHE2015} for more details).

On the other hand, the maximum mass of neutron star is still an open
question. There are some observational evidences which indicate that the
maximum mass of neutron stars can be more than $1.68M_{\odot }$. For
example, Jacoby et al \cite{Jacoby} and Verbiest et al \cite{Verbiest} for a
binary system used the detection of Shapiro delay to measure the masses of
both the neutron star and its binary companion. Also, using the same
approach, Demorest et al \cite{DemorestPRRH2010} has performed radio timing
observations for the binary millisecond pulsar PSR J1614-2230 where the
measured mass for this pulsar was obtained about $1.97M_{\odot }$. The mass
of other compact objects were obtained about $1.8M_{\odot }$ for Vela X-1
\cite{Rawls2011}, PSR J0348+0432 \cite{Antoniadis2013} about $2.01M_{\odot }$%
, 4U 1700-377 \cite{Clark2002} about $2.4M_{\odot }$, and J1748-2021B \cite%
{Freire2008} about $2.7M_{\odot }$. Also, A. W. Steiner et al determined an
empirical dense matter equation of state from a heterogeneous data set of
six neutron stars. They also showed that, the radius of a neutron star must
be in the range $R\leq \left( 11\thicksim 14\right) km$\ \cite{Steiner2010}.
In the present paper, we would like to see whether the gravity's rainbow and
obtained maximum mass for neutron stars by employing the modern equations of
state of neutron star matter derived from microscopic calculations can
predict existence of the maximum mass and radius of neutron stars more than $%
1.68M_{\odot }$\ and less than $14km$, respectively. Therefore, we consider
the neutron stars with radius less than $R\leq 14km$\ and separate the
maximum mass for $M_{\max }>2M_{\odot }$\ and $M_{\max }<2M_{\odot }$\ in
tables \ref{tab1} and \ref{tab12a}, respectively.
\begin{table}[tbp]
\caption{Structure properties of neutron star for different values of $H(%
\protect\varepsilon )$ ($1.2\leq $$H(\protect\varepsilon )$$\leq 1.67$).}
\label{tab1}
\begin{center}
\begin{tabular}{ccccccc}
\hline\hline
$H(\varepsilon)$ & ${M_{max}}\ (M_{\odot})$ & $R\ (km)$ & $R_{Sch}\ (km)$ & $%
\overline{\rho }$ $(10^{15}g$ $cm^{-3})$ & $\sigma (10^{-1})$ & $z(10^{-1})$
\\ \hline\hline
$1.70$ & $2.87$ & $14.32$ & $8.46$ & $0.46$ & $5.91$ & $5.63$ \\ \hline
$1.67$ & $2.81$ & $14.00$ & $8.28$ & $0.49$ & $5.91$ & $5.65$ \\ \hline
$1.60$ & $2.69$ & $13.47$ & $7.93$ & $0.52$ & $5.89$ & $5.59$ \\ \hline
$1.50$ & $2.52$ & $12.63$ & $7.43$ & $0.59$ & $5.88$ & $5.58$ \\ \hline
$1.40$ & $2.37$ & $11.79$ & $6.99$ & $0.69$ & $5.93$ & $5.66$ \\ \hline
$1.30$ & $2.19$ & $10.95$ & $6.46$ & $0.79$ & $5.90$ & $5.61$ \\ \hline
$1.20$ & $2.02$ & $10.10$ & $5.95$ & $0.93$ & $5.89$ & $5.61$ \\ \hline\hline
&  &  &  &  &  &
\end{tabular}%
\end{center}
\end{table}

\begin{table}[tbp]
\caption{Structure properties of neutron star for different values of $H(%
\protect\varepsilon )$ ($0.5\leq$$H(\protect\varepsilon )$$\leq1.1$).}
\label{tab12a}
\begin{center}
\begin{tabular}{ccccccc}
\hline\hline
$H(\varepsilon)$ & ${M_{max}}\ (M_{\odot})$ & $R\ (km)$ & $R_{Sch}\ (km)$ & $%
\overline{\rho }$ $(10^{15}g$ $cm^{-3})$ & $\sigma (10^{-1})$ & $z(10^{-1})$
\\ \hline\hline
$1.10$ & $1.85$ & $9.26$ & $5.45$ & $1.11$ & $5.88$ & $5.59$ \\ \hline
$1.00$ & $1.68$ & $8.42$ & $4.95$ & $1.34$ & $5.88$ & $5.58$ \\ \hline
$0.90$ & $1.51$ & $7.58$ & $4.45$ & $1.65$ & $5.87$ & $5.56$ \\ \hline
$0.80$ & $1.35$ & $6.73$ & $3.98$ & $2.10$ & $5.91$ & $5.64$ \\ \hline
$0.70$ & $1.18$ & $5.89$ & $3.48$ & $2.74$ & $5.91$ & $5.62$ \\ \hline
$0.60$ & $1.01$ & $5.05$ & $2.98$ & $3.72$ & $5.90$ & $5.61$ \\ \hline
$0.50$ & $0.84$ & $4.21$ & $2.48$ & $5.35$ & $5.89$ & $5.58$ \\ \hline\hline
&  &  &  &  &  &
\end{tabular}%
\end{center}
\end{table}
\begin{figure}[tbp]
$%
\begin{array}{cc}
\epsfxsize=7cm \epsffile{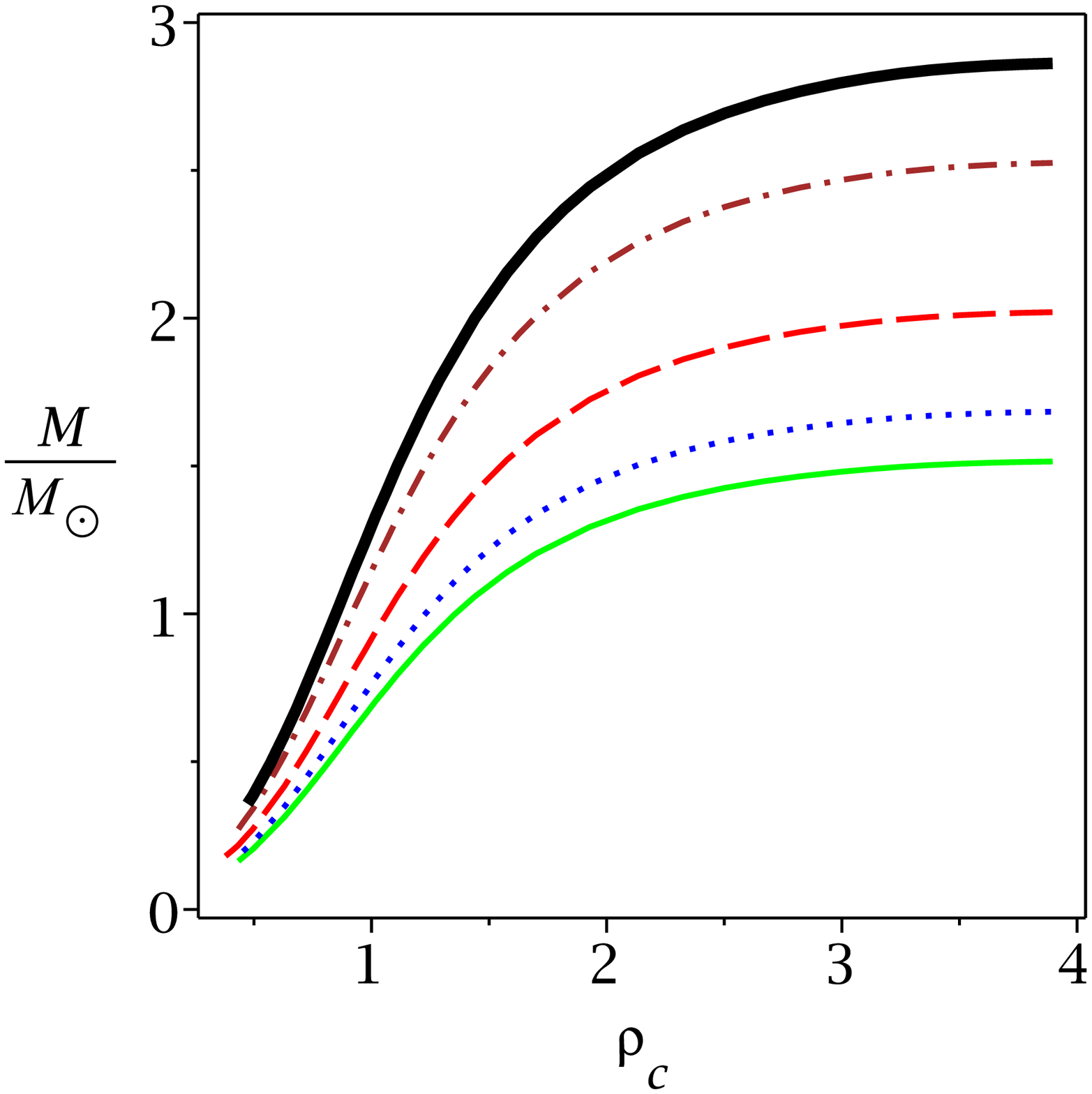} & \epsfxsize=7cm %
\epsffile{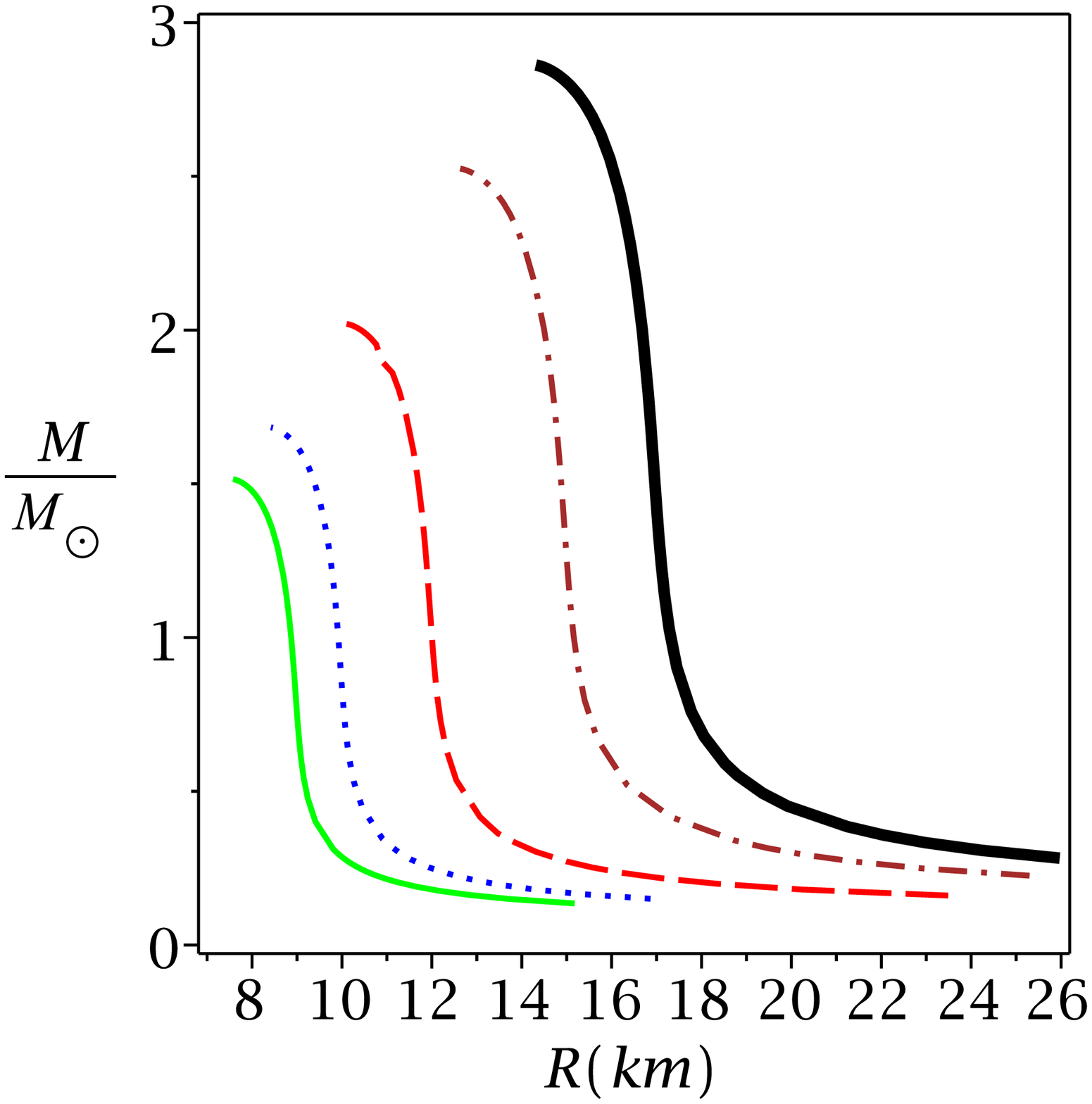}%
\end{array}
$%
\caption{Gravitational mass versus central mass density (left
diagram) and
radius (right diagram), for $H(\protect\varepsilon )=0.9$ (continuous line), $%
H(\protect\varepsilon )=1$ (doted line), $H(\protect\varepsilon )=1.2$
(dashed line), $H(\protect\varepsilon )=1.5$ (dashed-dotted line) and $H(%
\protect\varepsilon )=1.7$ (bold line).}
\label{Fig3}
\end{figure}

It is notable that, for $H(\varepsilon )=1$, the maximum mass of neutron
star reduces to the result that was obtained in Einstein gravity \cite%
{BordbarH2006,BordbarHE2015}, as expected (because, this case ($%
H(\varepsilon )=1)$ is denoted as the absence of gravity's rainbow. In other
words, in this case, the effects of the gravity's rainbow are vanished). On
the other hand, our results show that, by increasing rainbow function more
than $1$ ($H(\varepsilon )>1$), the maximum mass of neutron star increases
(see tables \ref{tab1} and \ref{tab12a}). In other words, our results cover
the mass measurement of massive neutron stars, and also, predict that the
mass of neutron stars in gravity's rainbow can be in the range larger than $%
1.68M_{\odot }$ (see tables \ref{tab1} and \ref{tab12a} for more details).
In order to conduct further investigations, we plot diagram related to the
gravitational mass versus central mass density (or radius) in Fig. \ref{Fig3}%
. The variation of the gravitational mass versus different values of the
rainbow function is presented in this figure. On the contrary, for $%
H(\varepsilon )<1$, both maximum mass of neutron stars and radius of these
stars are smaller than the Einstein case (see table \ref{tab12a} and Fig. %
\ref{Fig3}). This emphasizes the contributions of the gravity's rainbow on
properties of neutron stars.

In the above tables, we used the obtained TOV equations in Eqs. (\ref{TOVI})
and (\ref{TOVII}) for $H(\varepsilon )<1$\ and $H(\varepsilon )>1$,
respectively, to extract the structure of neutron stars. As one can see, due
to the structure of obtained TOV equation (Eq. (\ref{TOV})), for case (\ref%
{caseII}), the obtained TOV equation is independent of energy functions
which is not of our interest.

In the following, we are going to investigate other properties of neutron
star in the gravity's rainbow.

\subsubsection{Schwarzschild radius}

Using Eq. (\ref{4g(r)}), and this fact that $g(r=R_{Sch})=0$, we calculate
Schwarzschild radius for the obtained masses in gravity's rainbow as%
\begin{equation}
R_{Sch}=\frac{2GM_{eff}}{c^{2}}.  \label{Sch}
\end{equation}

As it was pointed out, the mass of neutron star depends on rainbow function
and thus we expect the Schwarzschild radius varies depending on choices of
rainbow function. The results are presented in tables \ref{tab1} and \ref%
{tab12a}. It is evident that, the Schwarzschild radius is an increasing
function of rainbow function.

\subsubsection{Average density}

The average density of a neutron star has the following form%
\begin{equation}
\overline{\rho }=\frac{3M_{eff}}{4\pi R^{3}},  \label{density}
\end{equation}%
where it is a decreasing function of $H(\varepsilon )$.

Obtained central density by using the mentioned equation of state of neutron
star matter is about $3.9\times 10^{15}\ g~cm^{-3}$ and this density is
larger than the normal nuclear density, $\rho _{0}=2.7\times 10^{14}g~cm^{-3}
$ \cite{WiringaFF1988}. On the other hand, by considering the rainbow
function less than $0.6$ ($H(\varepsilon )<0.6$), obtained average density
for neutron stars is larger than the central density (see the last row of
table \ref{tab12a}). In other words, when the rainbow function is larger
than $0.6$ ($H(\varepsilon )>0.6$), the neutron star is compatible.

\subsubsection{Compactness}

The compactness of a spherical object may be defined as the ratio of
Schwarzschild radius to the radius of object ($\sigma =R_{Sch}/R$), which
may be indicated as the strength of gravity. Obtained $\sigma $\ in this
gravity shows that the strength of gravity is almost the same for these
neutron stars.

\subsubsection{Gravitational redshift}

\begin{figure}[tbp]
$%
\begin{array}{c}
\epsfxsize=7cm \epsffile{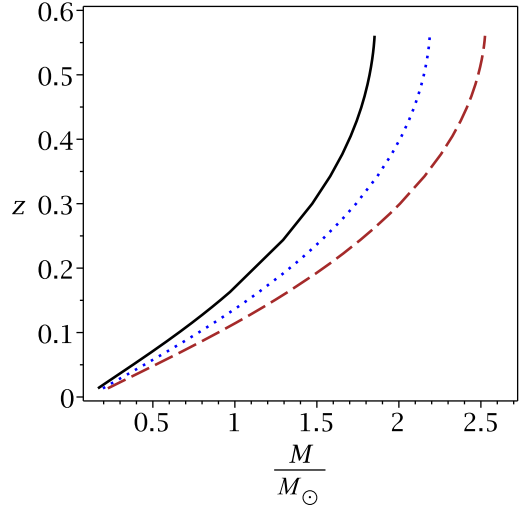}%
\end{array}
$%
\caption{The gravitational redshift versus mass for $H(\protect\varepsilon %
)=1.1$ (continuous line), $H(\protect\varepsilon )=1.3$ (doted line) and $H(%
\protect\varepsilon )=1.5$ (dashed line).}
\label{Figred}
\end{figure}

Using Eq. (\ref{4g(r)}), we can obtain the gravitational redshift in the
following form
\begin{equation}
z=\frac{1}{\sqrt{1-\frac{2GM_{eff}}{c^{2}R}}}-1,
\end{equation}%
where $M_{eff}$ is the mass of neutron stars in which is rainbow function
dependant. This relation indicates that to have a physical redshift, the
radius of a neutron star should be greater than the Schwarzschild radius, $R>%
\frac{2GM_{eff}}{c^{2}}=R_{Sch}$.\textbf{\ }The results related to the
gravitational redshift shows that for neutron stars with different mass,
this quantity is almost the same (see the last column of tables \ref{tab1}
and \ref{tab12a}). We plot the diagrams related to the gravitational
redshift versus mass in Fig. \ref{Figred}. This figure shows that, for
different values of rainbow function, this quantity is not affected
considerably.


\subsubsection{Dynamical stability}

The dynamical stability of the stellar model against the infinitesimal
radial adiabatic perturbation was introduced by Chandrasekhar \cite%
{Chandrasekhar}. Then, this stability condition was developed and applied to
astrophysical cases by many authors \cite%
{Bardeen,Kuntsem,MakH2013,KalamHM2015}. The adiabatic index ($\gamma $) is
defined in following form%
\begin{equation}
\gamma =\frac{\rho c^{2}+P}{c^{2}P}\frac{dP}{d\rho }.  \label{adiabatic}
\end{equation}

\begin{figure}[tbp]
$%
\begin{array}{c}
\epsfxsize=12cm \epsffile{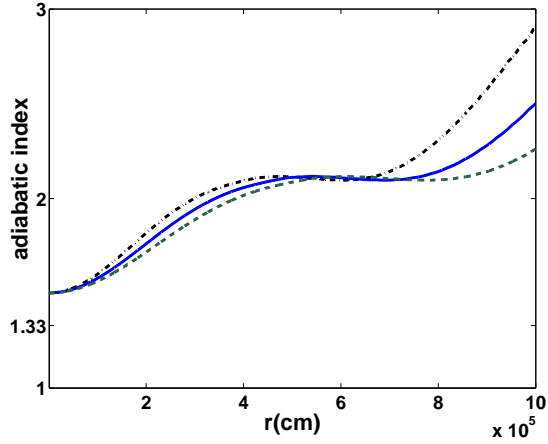}%
\end{array}
$%
\caption{Adiabatic index versus radius for $H(\protect\varepsilon
)=1.80$
(dashed line), $H(\protect\varepsilon )=1.60$ (continuous line) and $H(%
\protect\varepsilon )=1.4$ (dashed-dotted line).}
\label{Fig4}
\end{figure}

\begin{figure}[tbp]
$%
\begin{array}{c}
\epsfxsize=12cm \epsffile{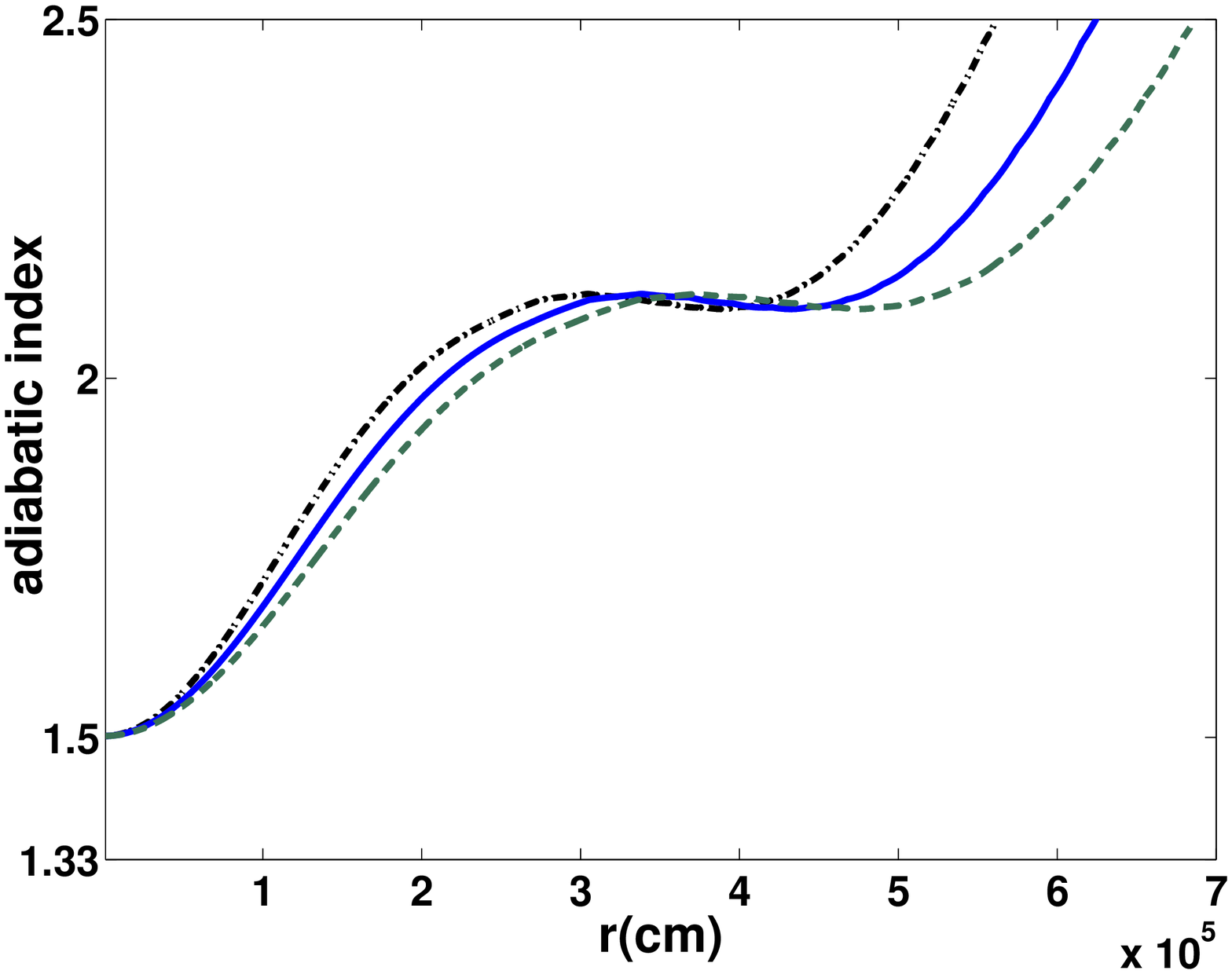}%
\end{array}
$%
\caption{Adiabatic index versus radius for $H(\protect\varepsilon
)=1.1$
(dashed line), $H(\protect\varepsilon )=1.0$ (continuous line) and $H(%
\protect\varepsilon )=0.9$ (dashed-dotted line).}
\label{Fig4b}
\end{figure}

In order to investigate the dynamical stability condition, $\gamma $ should
be more than $\frac{4}{3}$ ($\gamma >\frac{4}{3}=1.33$) everywhere within
the isotropic star. For this purpose, we plot two diagrams related to $%
\gamma $ versus radius for different values of rainbow functions in Figs. %
\ref{Fig4} and \ref{Fig4b}. As one can see, this stellar model is stable
against the radial adiabatic infinitesimal perturbations.

\begin{figure}[tbp]
$%
\begin{array}{c}
\epsfxsize=12cm \epsffile{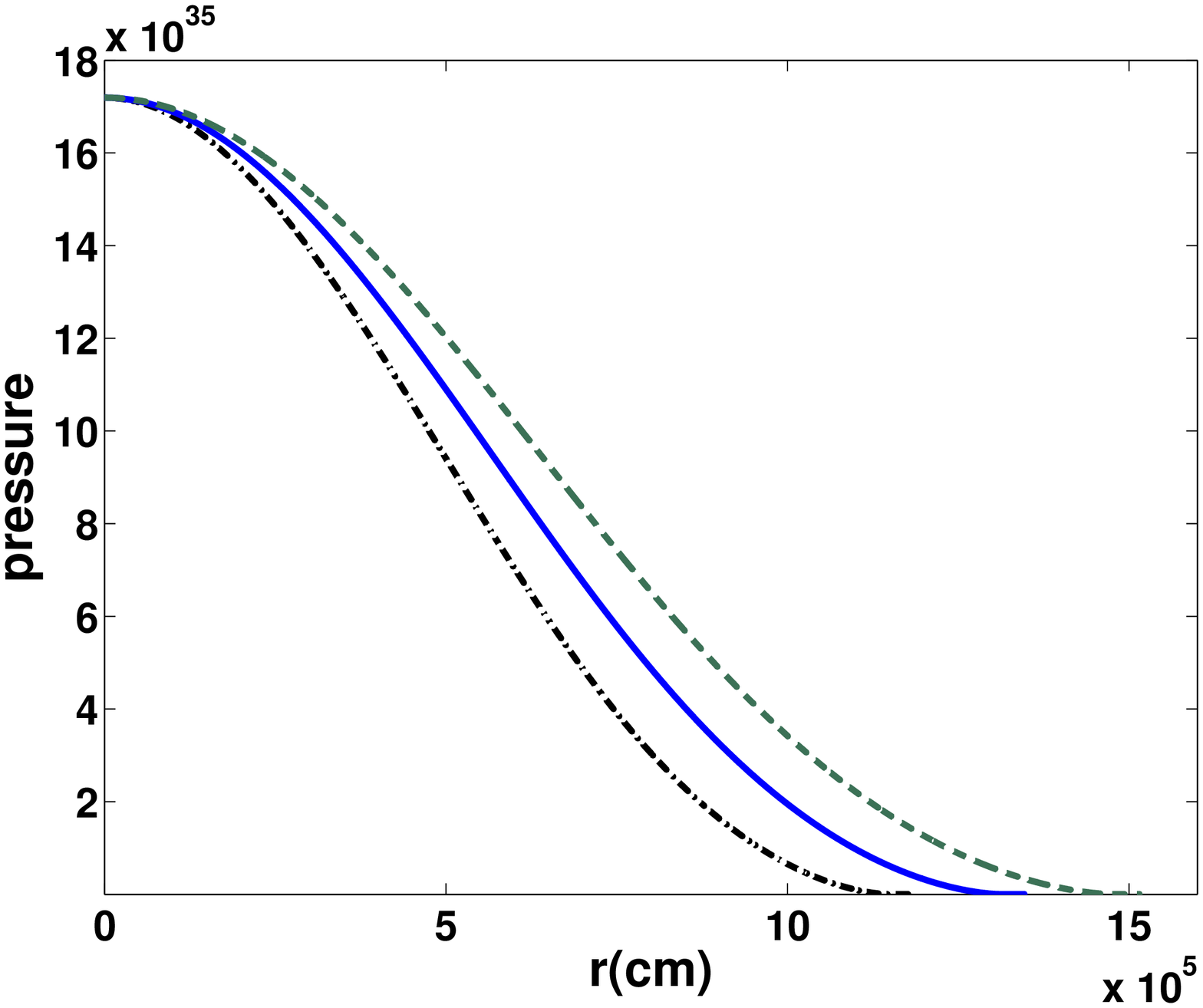}%
\end{array}
$%
\caption{Pressure (erg/$cm^{3}$) versus radius for $H(\protect\varepsilon %
)=1.80$ (dashed line), $H(\protect\varepsilon )=1.60$ (continuous line) and $%
H(\protect\varepsilon )=1.4$ (dashed-dotted line).}
\label{Fig5}
\end{figure}

\begin{figure}[tbp]
$%
\begin{array}{c}
\epsfxsize=12cm \epsffile{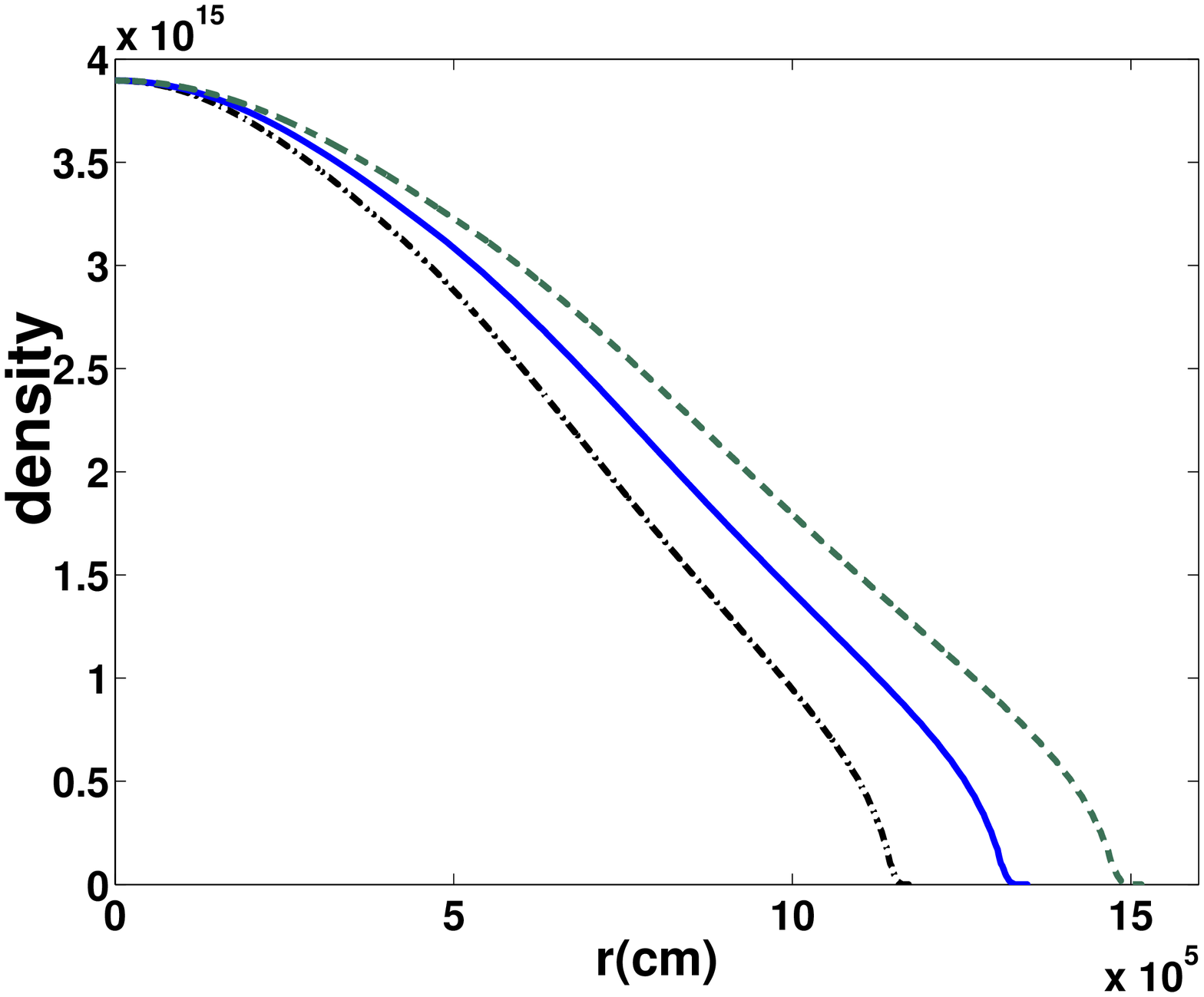}%
\end{array}
$%
\caption{Density (gr/$cm^{3}$) versus radius for $H(\protect\varepsilon %
)=1.80$ (dashed line), $H(\protect\varepsilon )=1.60$ (continuous line) and $%
H(\protect\varepsilon )=1.4$ (dashed-dotted line).}
\label{Fig6}
\end{figure}

Also, we plot the pressure (density) versus distance from the center of
neutron star. As one can see, the pressure and density are maximum at the
center and decreases monotonically towards the boundary (see Figs \ref{Fig5}
and \ref{Fig6}).

As one can see, considering the radius of neutron stars less than $14km$
(according to the work of A. W. Steiner et al. \cite{Steiner2010}) and by
using the average density, we obtained an upper and a lower limitations for
rainbow function. Therefore, the obtained solutions in gravity's rainbow are
valid in the range of $0.6<H(\varepsilon )\leq 1.67$. In other words,
gravity's rainbow cover all neutron stars and pulsars which have the maximum
mass less than $2.81M_{\odot }$ (see tables \ref{tab1} and \ref{tab12a}).
Indeed, these results cover wide range of neutron stars and pulsars such as
Vela X-1, PSR J0348+0432, 4U 1700-377, J1748-2021B, PSR J1903 + 327, Cen X -
3, PSR B1913 + 16, PSR J0737 - 3039, PSR J0737 - 3039B and SMC X - 1.

\subsection{\textbf{Relation between the mass of neutron stars and Planck
mass}}

Here we derive the relation between neutron star mass and the Planck mass in
the gravity's rainbow according to Burrows and Ostriker's work \cite{Burrows}%
. We know that in addition to the degenerate pressure of nucleons, the
pressure due to the strong repulsive inter-nucleons force balances the
pressure due to the gravitation force in the neutron stars. Here, the
nucleon-nucleon interaction which is so strong, is taken place through the
pion exchange. Using this idea, we can consider the average density of a
neutron star in term of the nucleus density using the following relation
\begin{equation}
\rho _{nuc}\sim \frac{3m_{p}}{4\pi \lambda _{\pi }^{3}},
\end{equation}%
where $m_{p}$\ is the proton mass and $\lambda _{\pi }=\frac{\hbar }{m_{\pi
}c}$ is the Compton wavelength. Here $m_{\pi }$\ is the pion mass. On the
other hand, the mass of a neutron star is set by a general relativity
instability. The condition for Einstein gravity instability is $R=\frac{2GM}{%
c^{2}}$\ (see Ref. \cite{Burrows} for more details). Now, we use this idea
for gravity's rainbow. The Schwarzschild radius and the average density of a
neutron star in gravity's rainbow are obtained by equations (\ref{Sch}) and (%
\ref{density}) as follows
\begin{eqnarray}
R_{Sch} &=&\frac{2GM_{eff}}{c^{2}}=\frac{2GM}{c^{2}H^{2}(\varepsilon )},
\label{RschI} \\
&&  \notag \\
\overline{\rho } &=&\frac{3M_{eff}}{4\pi R^{3}}=\frac{3M}{4\pi
H^{2}(\varepsilon )R^{3}},  \label{Den}
\end{eqnarray}%
where\textbf{\ }
\begin{equation}
M_{eff}\left( r,\varepsilon \right) =\frac{1}{H^{2}(\varepsilon )}\int 4\pi
r^{2}\rho (r)dr.
\end{equation}%
Therefore, we have $M_{eff}=\frac{M}{H^{2}(\varepsilon )}$. Now, we can
derive the corresponding mass using the equations (\ref{RschI}) and (\ref%
{Den}) and considering $\rho _{nuc}$ as follows
\begin{eqnarray}
M &\sim &H^{2}(\varepsilon )\left( \frac{\hbar c}{G}\right) ^{3/2}\frac{1}{%
m_{p}^{2}}\left( \frac{\eta _{\pi }}{2\eta _{p}}\right) ^{3/2}  \notag \\
&\sim &H^{2}(\varepsilon )M_{Ch}\left( \frac{\eta _{\pi }}{2\eta _{p}}%
\right) ^{3/2}  \notag \\
&\sim &H^{2}(\varepsilon )m_{pl}\eta _{p}^{2}\left( \frac{\eta _{\pi }}{%
2\eta _{p}}\right) ^{3/2},  \label{Planck}
\end{eqnarray}%
where $M_{Ch}$\ is the Chandrasekhar mass. In Eq. (\ref{Planck}), $%
M_{Ch}\sim $\ $\left( \frac{\hbar c}{G}\right) ^{3/2}\frac{1}{m_{p}^{2}}\sim
m_{pl}\eta _{p}^{2}$, $\eta _{p}=\frac{m_{pl}}{m_{p}}$ and $\eta _{\pi }=%
\frac{m_{pl}}{m_{\pi }}$ where $m_{pl}$\ is the Planck mass \ (see Ref. \cite%
{Burrows} for more details).  \ Therefore, the relation between the neutron
star mass and the Planck mass is obtained by above equation. It is notable
that for $H(\varepsilon )=1$, Eq. (\ref{Planck}) reduces to Einstein gravity
\cite{Burrows}. On the other hand, we can rewrite the equation (\ref{Planck}%
) in the following form
\begin{equation}
M\sim H^{2}(\varepsilon )\Psi _{EN},  \label{Limit}
\end{equation}%
where $\Psi _{EN}=m_{pl}\eta _{p}^{2}\left( \frac{\eta _{\pi }}{2\eta _{p}}%
\right) ^{3/2}$\textbf{.}

\subsection{Properties of neutron stars in gravity's rainbow with
cosmological constant}

In this subsection, we take the effects of variation of the cosmological
constant into account. In other words, the effects of variation of the
cosmological constant on maximum mass and radius of the neutron stars in the
presence of Einstein gravity's rainbow are investigated. Our results are
presented in table \ref{tab2}.
\begin{table}[tbp]
\caption{Structure properties of neutron star for $H(\protect\varepsilon )=1$
(up table) and $H(\protect\varepsilon )=1.5$ (down table), respectively,
with different values of $\Lambda $. }
\label{tab2}
\begin{center}
\begin{tabular}{c}
\begin{tabular}{ccccccc}
\hline\hline
$\Lambda\ (m^{-2})$ & ${M_{max}}\ (M_{\odot})$ & $R\ (km)$ & $R_{Sch}\ (km)$
& $\overline{ \rho }$ $(10^{15}g$ $cm^{-3})$ & $\sigma (10^{-1})$ & $%
z(10^{-1})$ \\ \hline\hline $1.00\times 10^{-16}$ & $1.68$ &
$8.42$ & $4.95$ & $1.34$ & $5.88$ & $5.58$
\\ \hline
$1.00\times 10^{-14}$ & $1.68$ & $8.42$ & $4.95$ & $1.34$ & $5.88$
& $5.58$
\\ \hline
$5.00\times 10^{-14}$ & $1.68$ & $8.41$ & $4.95$ & $1.33$ & $5.88$ & $5.58$
\\ \hline
$1.00\times 10^{-13}$ & $1.67$ & $8.40$ & $4.92$ & $1.33$ & $5.86$ & $5.54$
\\ \hline
$5.00\times 10^{-13}$ & $1.62$ & $8.34$ & $4.77$ & $1.32$ & $5.72$ & $5.29$
\\ \hline
$1.00\times 10^{-12}$ & $1.56$ & $8.25$ & $4.60$ & $1.31$ & $5.57$ & $5.03$
\\ \hline
$5.00\times 10^{-12}$ & $1.12$ & $7.47$ & $3.30$ & $1.27$ & $4.42$ & $3.38$
\\ \hline
$1.00\times 10^{-11}$ & $0.78$ & $6.65$ & $2.29$ & $1.25$ & $3.46$ & $2.36$
\\ \hline\hline
&  &  &  &  &  &
\end{tabular}
\\
\begin{tabular}{ccccccc}
\hline\hline
$\Lambda\ (m^{-2})$ & ${M_{max}}\ (M_{\odot})$ & $R\ (km)$ & $R_{Sch}\ (km)$
& $\overline{\rho }$ $(10^{14}g$ $cm^{-3})$ & $\sigma (10^{-1})$ & $%
z(10^{-1})$ \\ \hline\hline
$1.00\times 10^{-16}$ & $2.52$ & $12.63$ & $7.43$ & $5.94$ & $5.88$ & $5.58$
\\ \hline
$1.00\times 10^{-14}$ & $2.52$ & $12.63$ & $7.43$ & $5.94$ & $5.88$ & $5.58$
\\ \hline
$5.00\times 10^{-14}$ & $2.51$ & $12.62$ & $7.40$ & $5.94$ & $5.86$ & $5.54$
\\ \hline
$1.00\times 10^{-13}$ & $2.50$ & $12.61$ & $7.37$ & $5.92$ & $5.85$ & $5.51$
\\ \hline
$5.00\times 10^{-13}$ & $2.44$ & $12.51$ & $7.19$ & $5.91$ & $5.75$ & $5.34$
\\ \hline
$1.00\times 10^{-12}$ & $2.34$ & $12.38$ & $6.90$ & $5.86$ & $5.57$ & $5.03$
\\ \hline
$5.00\times 10^{-12}$ & $1.68$ & $11.21$ & $4.95$ & $5.69$ & $4.41$ & $3.38$
\\ \hline
$1.00\times 10^{-11}$ & $1.17$ & $9.98$ & $3.45$ & $5.58$ & $3.46$ & $2.36$
\\ \hline\hline
&  &  &  &  &  &
\end{tabular}%
\end{tabular}%
\end{center}
\end{table}
\begin{table}[tbp]
\caption{Properties of neutron star with and without the cosmological
constant.}
\label{tab22}
\begin{center}
\begin{tabular}{cccccccc}
\hline\hline
$H(\varepsilon)$ & $\Lambda\ (m^{-2})$ & ${M_{max}}\ (M_{\odot})$ & $R\ (km)$
& $R_{Sch}\ (km)$ & $\overline{\rho }$ $(10^{15}g$ $cm^{-3})$ & $\sigma
(10^{-1})$ & $z(10^{-1})$ \\ \hline\hline
$1.00$ & $%
\begin{array}{c}
1\times 10^{-52} \\
0%
\end{array}
$ & $%
\begin{array}{c}
1.68 \\
1.68%
\end{array}
$ & $%
\begin{array}{c}
8.42 \\
8.42%
\end{array}
$ & $%
\begin{array}{c}
4.95 \\
4.95%
\end{array}
$ & $%
\begin{array}{c}
1.34 \\
1.34%
\end{array}
$ & $%
\begin{array}{c}
5.88 \\
5.88%
\end{array}
$ & $%
\begin{array}{c}
5.58 \\
5.58%
\end{array}
$ \\ \hline
$1.30$ & $%
\begin{array}{c}
1\times 10^{-52} \\
0%
\end{array}
$ & $%
\begin{array}{c}
2.19 \\
2.19%
\end{array}
$ & $%
\begin{array}{c}
10.95 \\
10.95%
\end{array}
$ & $%
\begin{array}{c}
6.46 \\
6.46%
\end{array}
$ & $%
\begin{array}{c}
0.79 \\
0.79%
\end{array}
$ & $%
\begin{array}{c}
5.90 \\
5.90%
\end{array}
$ & $%
\begin{array}{c}
5.61 \\
5.61%
\end{array}
$ \\ \hline
$1.60$ & $%
\begin{array}{c}
1\times 10^{-52} \\
0%
\end{array}
$ & $%
\begin{array}{c}
2.69 \\
2.69%
\end{array}
$ & $%
\begin{array}{c}
13.47 \\
13.47%
\end{array}
$ & $%
\begin{array}{c}
7.93 \\
7.93%
\end{array}
$ & $%
\begin{array}{c}
0.52 \\
0.52%
\end{array}
$ & $%
\begin{array}{c}
5.89 \\
5.89%
\end{array}
$ & $%
\begin{array}{c}
5.59 \\
5.59%
\end{array}
$ \\ \hline\hline
&  &  &  &  &  &  &
\end{tabular}%
\end{center}
\end{table}

\begin{figure}[tbp]
$%
\begin{array}{c}
\epsfxsize=7cm \epsffile{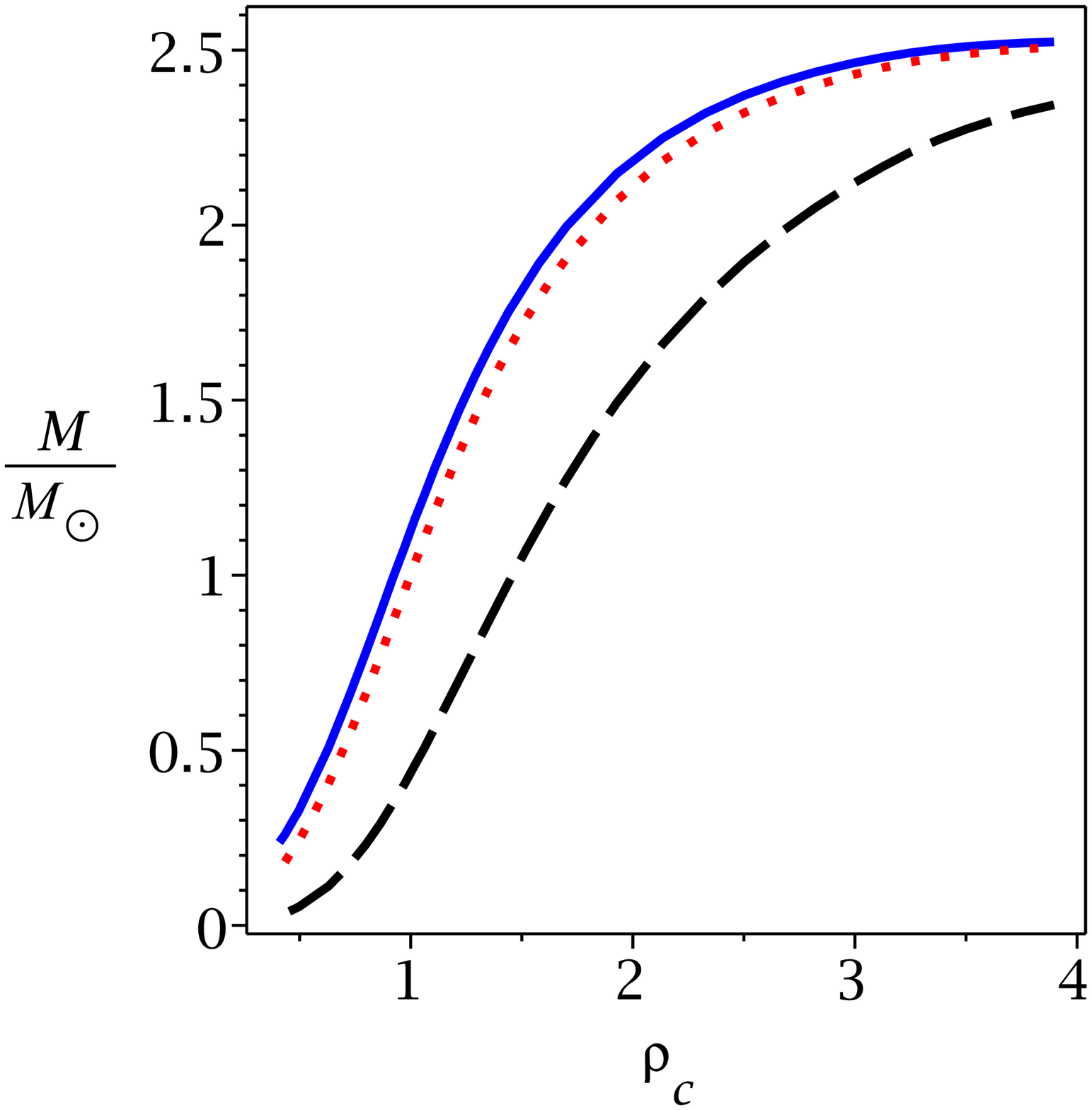}%
\end{array}
$%
\caption{Gravitational mass of neutron star versus central mass density for $%
H(\protect\varepsilon)=1.5$, $\Lambda=1\times 10^{-14}$ (continuous line), $%
\Lambda=1\times 10^{-13}$ (doted line) and $\Lambda=1\times 10^{-12}$
(dashed line).}
\label{Fig7}
\end{figure}

As one can see, interestingly, the maximum mass of neutron star is a
decreasing function of the cosmological constant (see table \ref{tab2} for
more details). For further investigations of the behavior of the mass versus
radius, we have plotted the following diagrams (Figs. \ref{Fig7} and \ref%
{Fig8}).

\begin{figure}[tbp]
$%
\begin{array}{ccc}
\epsfxsize=5cm \epsffile{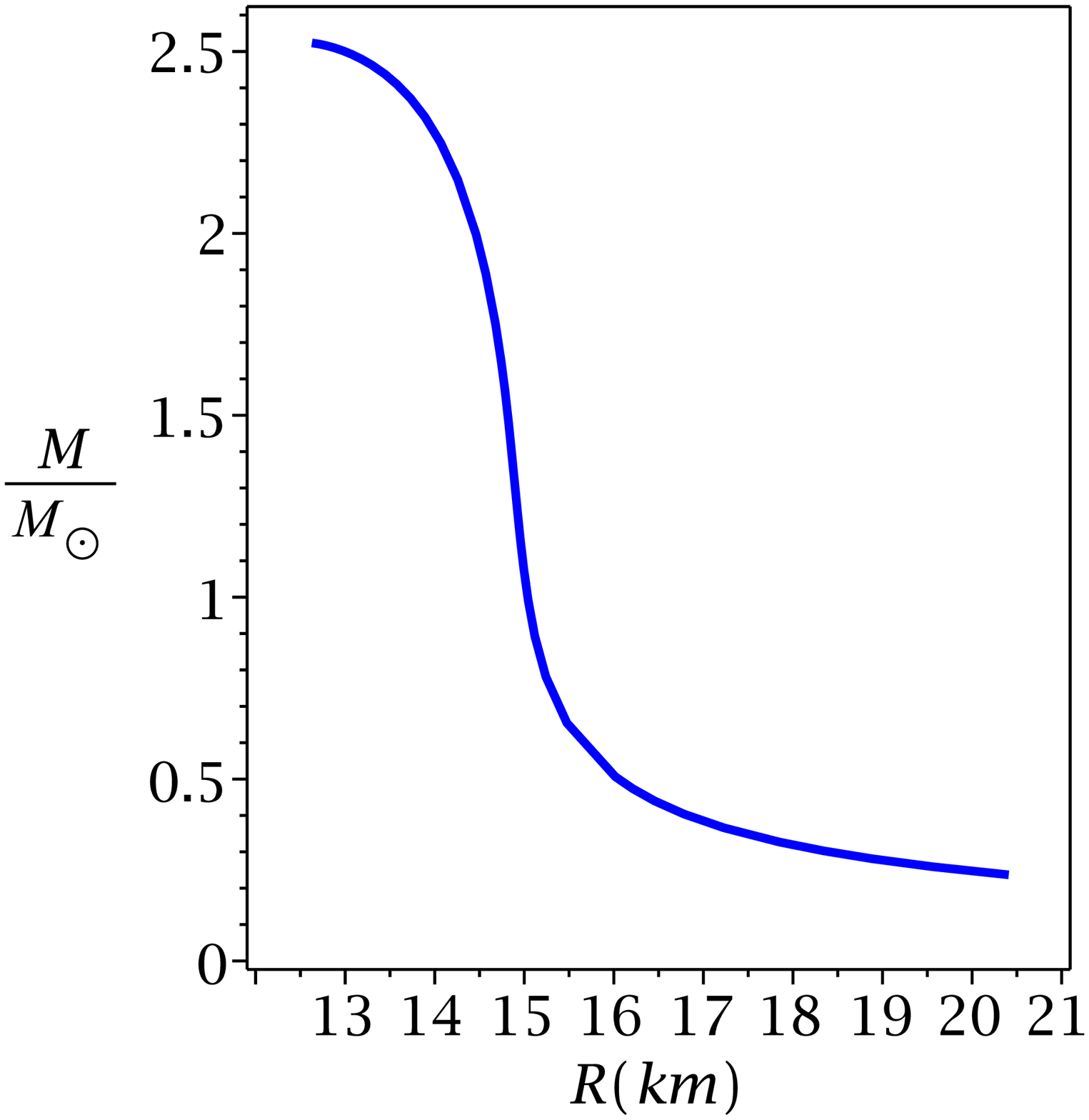} & \epsfxsize=5cm %
\epsffile{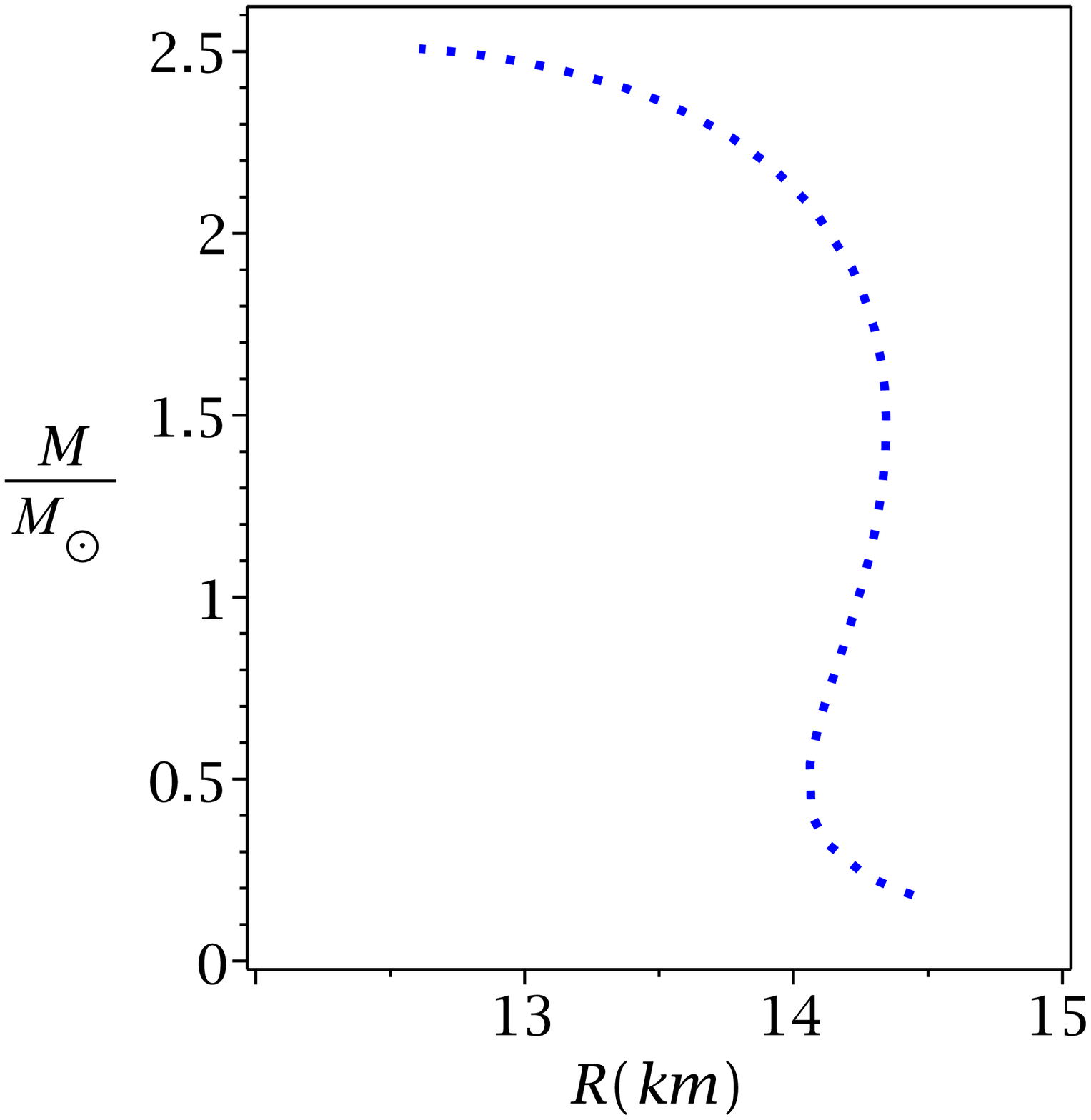} & \epsfxsize=5cm \epsffile{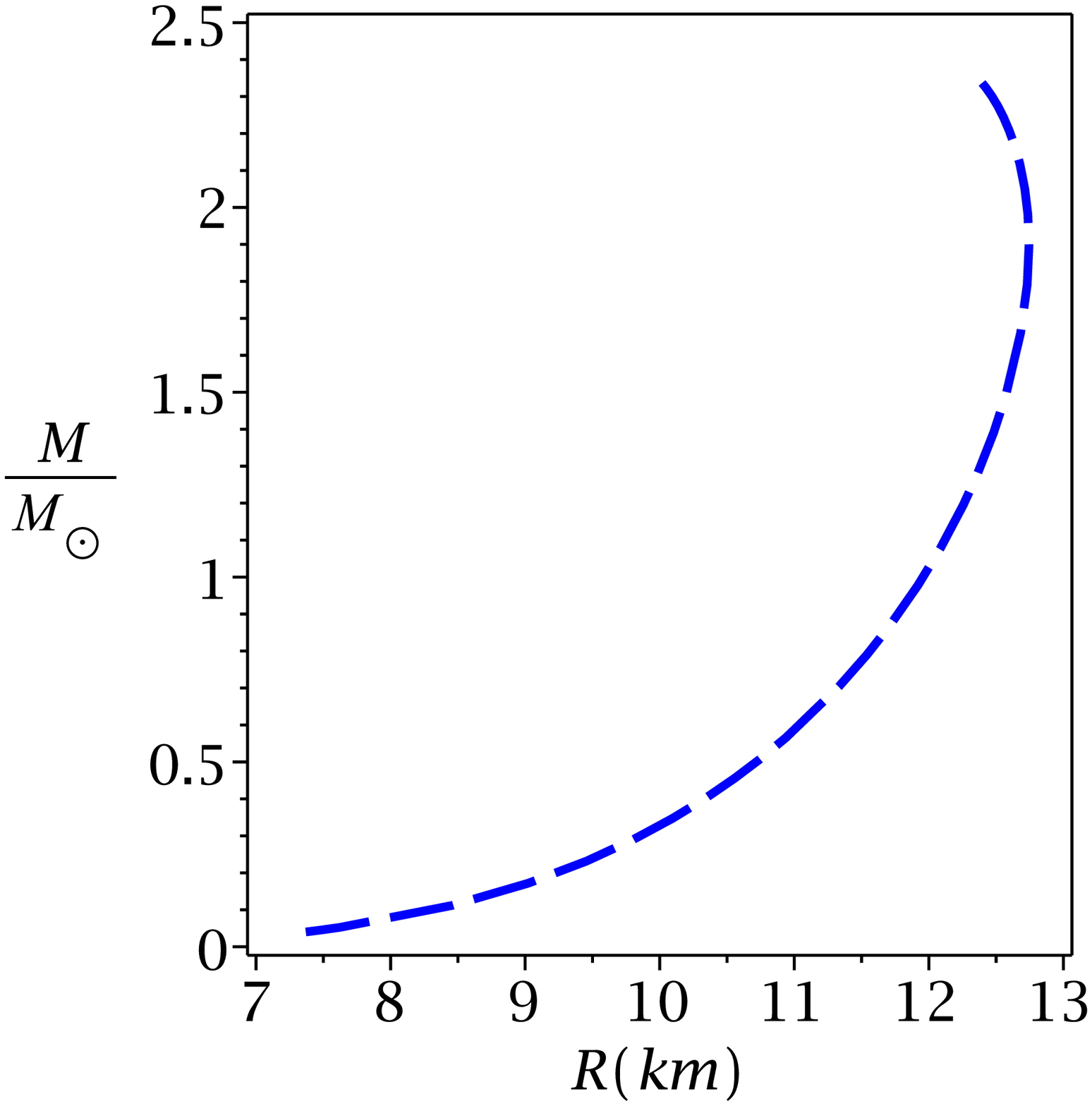}%
\end{array}
$%
\caption{Gravitational mass of neutron star versus radius for $H(\protect%
\varepsilon)=1.5$, $\Lambda=1\times 10^{-14}$, $\Lambda=1\times
10^{-13}$ and $\Lambda=1\times 10^{-12}$ , from left to right,
respectively.} \label{Fig8}
\end{figure}

It is evident from Figs. \ref{Fig7} and \ref{Fig8} that the behavior of mass
as a function of the central mass density (or radius) is highly sensitive to
the variation of cosmological constant. For large values of cosmological
constat (Fig. \ref{Fig8}), the behavior is completely modified. In this
case, in the presence of large values of cosmological constant, the behavior
of the mass versus radius of neutron star will be modified into a quark like
behavior.

It is a matter of calculation to show that the Schwarzschild radius of
gravity's rainbow in the presence of the cosmological constant is%
\begin{equation}
R_{Sch}=\frac{\left[ H^{2}(\varepsilon )\left( \frac{3GM_{eff}}{c^{2}}+\sqrt{%
\frac{H^{2}(\varepsilon )}{\Lambda }+\frac{9G^{2}M_{eff}^{2}}{c^{4}}}\right) %
\right] ^{1/3}}{\Lambda ^{1/3}}-\frac{H^{4/3}(\varepsilon )}{\left[ \left(
\frac{3GM_{eff}}{c^{2}}+\sqrt{\frac{H^{2}(\varepsilon )}{\Lambda }+\frac{%
9G^{2}M_{eff}^{2}}{c^{4}}}\right) \Lambda ^{2}\right] ^{1/3}}.  \notag
\end{equation}

The above equation shows that the Schwarzschild radius is modified in the
gravity's rainbow. In other words, the Schwarzschild radius depends on the
rainbow function. It is notable that, when $H(\varepsilon )=1$, the
Schwarzschild radius reduces to the obtained Schwarzschild radius in Ref.
\cite{BordbarHE2015}. The results show that the Schwarzschild radius is a
decreasing function of the cosmological constant (see table \ref{tab2} for
more details). We present the results of the average density and gravity
strength in the presence of the cosmological constant in table \ref{tab2}.
We find that increasing the cosmological constant leads to decreasing both
the strength of gravity and the maximum mass of neutron star. On the other
hand, calculations related to the average density show that by increasing $%
\Lambda $, the average density decreases.

We extract the gravitational redshift in gravity's rainbow and in the
presence of cosmological constant in the following form
\begin{equation}
z=\frac{1}{\sqrt{1+\frac{\Lambda R^{2}}{3H^{2}(\varepsilon )}-\frac{2GM_{eff}%
}{c^{2}R}}}-1.
\end{equation}

The results show that, the gravitational redshift is a decreasing function
of the cosmological constant and an increasing function of $M_{eff}$. In
addition, in order to have a finite and real redshift, we should set $1+%
\frac{\Lambda R^{2}}{3H^{2}(\varepsilon )}-\frac{2GM_{eff}}{c^{2}R}>0$ and
therefore, $R>R_{Sch}$, as we expected.

\begin{figure}[tbp]
$%
\begin{array}{c}
\epsfxsize=12cm \epsffile{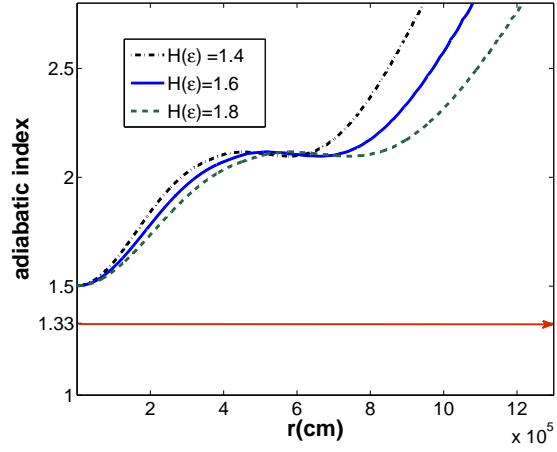}%
\end{array}
$%
\caption{Adiabatic index versus radius for $\Lambda =1\times 10^{-12}$, $H(%
\protect\varepsilon )=1.80$ (dashed line), $H(\protect\varepsilon )=1.60$
(continuous line) and $H(\protect\varepsilon )=1.40$ (dashed-dotted line).}
\label{Fig9}
\end{figure}

\begin{figure}[tbp]
$%
\begin{array}{c}
\epsfxsize=12cm \epsffile{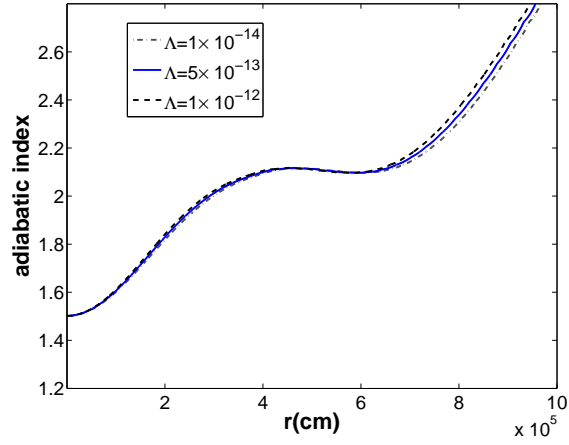}%
\end{array}
$%
\caption{Adiabatic index versus radius for $H(\protect\varepsilon )=1.40$, $%
\Lambda =1\times 10^{-12}$ (dashed line), $\Lambda =5\times 10^{-13}$
(continuous line) and $\Lambda =1\times 10^{-14}$ (dashed-dotted line).}
\label{Fig12}
\end{figure}

Also, the dynamical stability for neutron stars in gravity's rainbow and in
the presence of cosmological constant shows that these stars have dynamical
stability in all over the neutron stars (see Figs. \ref{Fig9} and \ref{Fig12}%
, for more details).

\begin{figure}[tbp]
$%
\begin{array}{c}
\epsfxsize=12cm \epsffile{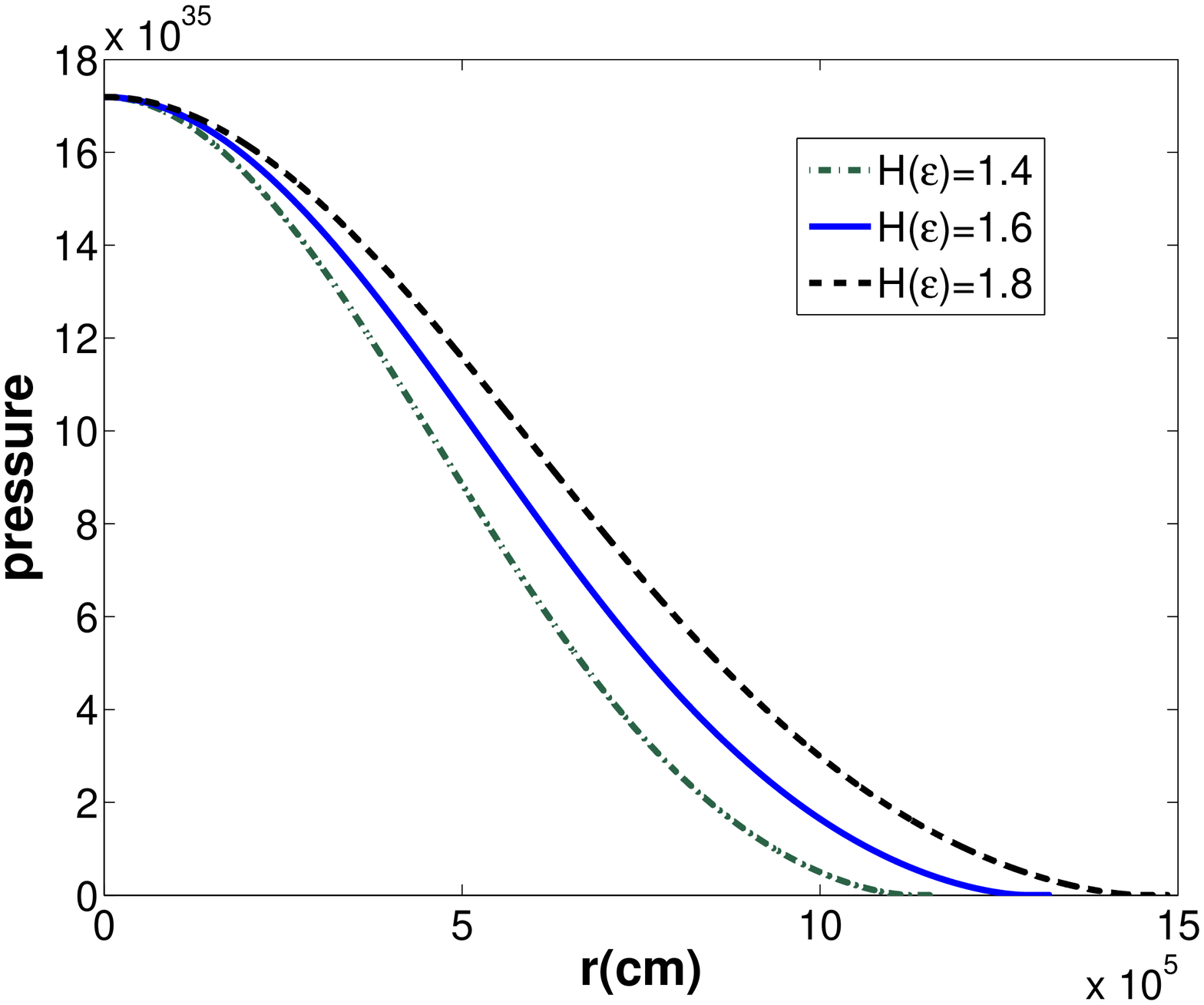}%
\end{array}
$%
\caption{Pressure (erg/$cm^{3}$) versus radius for $\Lambda
=1\times
10^{-12} $, $H(\protect\varepsilon )=1.80$ (dashed line), $H(\protect%
\varepsilon )=1.60$ (continuous line) and $H(\protect\varepsilon )=1.40$
(dashed-dotted line).}
\label{Fig10}
\end{figure}

\begin{figure}[tbp]
$%
\begin{array}{c}
\epsfxsize=12cm \epsffile{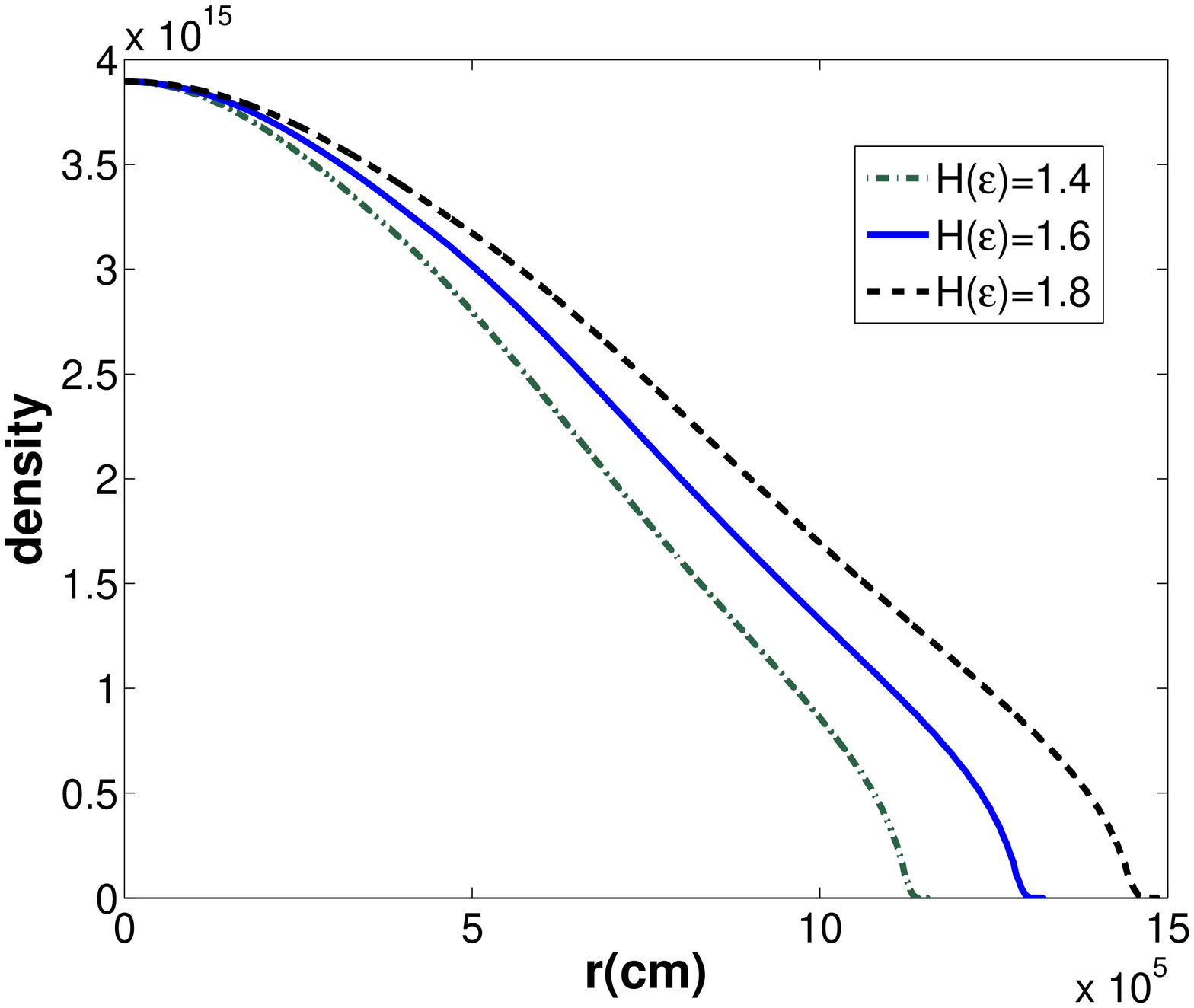}%
\end{array}
$%
\caption{Density (gr/$cm^{3}$) versus radius for $\Lambda =1\times
10^{-12}$
, $H(\protect\varepsilon )=1.80$ (dashed line), $H(\protect\varepsilon %
)=1.60 $ (continuous line) and $H(\protect\varepsilon )=1.40$ (dashed-dotted
line).}
\label{Fig11}
\end{figure}

In order to investigate internal structure of the neutron star in more
details, we plot the pressure (density) versus distance from the center of
neutron star in the presence the cosmological constant in gravity's rainbow.
Figures \ref{Fig10} and \ref{Fig11} show that, the pressure and density are
maximum at the center and decrease monotonically towards the boundary.

The value of the cosmological constant is an open question. On the other
hand, from the perspective of cosmology, this value is about $10^{-52}$ $%
m^{-2}$. As one can see, the cosmological constant has no significant effect
when its value is about $10^{-52}$ $m^{-2}$ (see table \ref{tab22} for more
details). In order to examine the effects of the cosmological constant on
properties of the neutron star, we should consider a toy model in which its
value is about less than $10^{-14}\ m^{-2}$. The results show that by
decreasing the cosmological constant (less than $10^{-14}$ $m^{-2}$, $%
\Lambda <10^{-14}\ m^{-2}$), the maximum mass and also the radius of this
star are not modified. In other words, for $\Lambda <10^{-14}\ m^{-2}$, the
cosmological constant does not affect the maximum and radius of neutron star
(see table \ref{tab2} for more details). One of the results of this work is
that when the value of the cosmological constant is about $10^{-52}~m^{-2}$,
this constant does not play a role in the structure of neutron stars, but by
taking larger values for it, the maximum mass and its radius are reduced.

\section{\textbf{Higher dimensions \label{Higher}}}

In this section, we extend our solutions to $5$\ and $6$\ dimensions to
obtain the maximum mass of neutron stars by using the mentioned equation of
state of neutron star matter.

First, we consider the TOV equation for gravity's rainbow without the
cosmological constant. The results of our calculations show that the maximum
mass of the neutron stars in $5$\ and $6$\ dimensions is not reasonable. In
other words, the obtained maximum mass of neutron star is about $%
10^{6}M_{\odot }$\ and $10^{12}M_{\odot }$\ for $5$\ and $6$\ dimensions,
respectively (see table \ref{tab5} for more details). In order to get a
better picture of the results, we calculate the Schwarzschild radius for the
obtained masses in gravity's rainbow at these dimensions. The Schwarzschild
radius for this gravity in higher dimensions is%
\begin{equation}
R_{Sch}=\left( \frac{8G_{d}M_{eff}\Gamma \left( \frac{d-1}{2}\right) }{%
c^{2}\left( d-2\right) \pi ^{\frac{d-3}{2}}}\right) ^{\frac{1}{d-3}}.
\end{equation}

Also, we discuss the average density of the neutron star in $5$\ and $6$\
dimensions, in which can be written as%
\begin{equation}
\overline{\rho }=\left\{
\begin{array}{cc}
\frac{2M_{eff}}{\pi ^{2}R^{4}}, & d=5 \\
\frac{15M_{eff}}{8\pi ^{2}R^{5}}, & d=6%
\end{array}%
\right. .
\end{equation}%
%
%
%
%
%
\begin{table}[tbp]
\caption{Properties of neutron star with and without the cosmological
constant.}
\label{tab5}
\begin{center}
\begin{tabular}{cccccccc}
\hline\hline
$H(\varepsilon)$ & $Dimensions$ & ${M_{max}}\ (M_{\odot})$ & $R\ (km)$ & $%
R_{Sch}\ (km)$ & $\overline{\rho }$ $(10^{15}g$ $cm^{-3})$ & $\sigma$ &  \\
\hline\hline
$1.70$ & $%
\begin{array}{c}
5 \\
6%
\end{array}
$ & $%
\begin{array}{c}
9.22\times 10^{6} \\
32.06\times 10^{12}%
\end{array}
$ & $%
\begin{array}{c}
17.52 \\
20.58%
\end{array}
$ & $%
\begin{array}{c}
107.41 \\
341.98%
\end{array}
$ & $%
\begin{array}{c}
0.39 \\
0.33%
\end{array}
$ & $%
\begin{array}{c}
6.13 \\
16.62%
\end{array}
$ &  \\ \hline
$1.40$ & $%
\begin{array}{c}
5 \\
6%
\end{array}
$ & $%
\begin{array}{c}
6.26\times 10^{6} \\
17.91\times 10^{12}%
\end{array}
$ & $%
\begin{array}{c}
14.43 \\
16.95%
\end{array}
$ & $%
\begin{array}{c}
88.50 \\
281.65%
\end{array}
$ & $%
\begin{array}{c}
0.58 \\
0.48%
\end{array}
$ & $%
\begin{array}{c}
6.13 \\
15.72%
\end{array}
$ &  \\ \hline
$1.10$ & $%
\begin{array}{c}
5 \\
6%
\end{array}
$ & $%
\begin{array}{c}
3.86\times 10^{6} \\
8.68\times 10^{12}%
\end{array}
$ & $%
\begin{array}{c}
11.33 \\
13.32%
\end{array}
$ & $%
\begin{array}{c}
69.09 \\
221.23%
\end{array}
$ & $%
\begin{array}{c}
0.94 \\
0.78%
\end{array}
$ & $%
\begin{array}{c}
6.10 \\
16.73%
\end{array}
$ &  \\ \hline
$0.80$ & $%
\begin{array}{c}
5 \\
6%
\end{array}
$ & $%
\begin{array}{c}
2.04\times 10^{6} \\
3.34\times 10^{12}%
\end{array}
$ & $%
\begin{array}{c}
8.24 \\
9.68%
\end{array}
$ & $%
\begin{array}{c}
50.52 \\
160.91%
\end{array}
$ & $%
\begin{array}{c}
1.78 \\
1.48%
\end{array}
$ & $%
\begin{array}{c}
6.13 \\
16.62%
\end{array}
$ &  \\ \hline\hline
&  &  &  &  &  &  &
\end{tabular}%
\end{center}
\end{table}
The results are collected in table \ref{tab5}. These results show that for $5
$\ and $6$\ dimensions, obtained radius of the neutron stars in gravity's
rainbow are within the Schwarzschild radius. In other words, these objects
can not be neutron stars, but black holes. So the mentioned equation of
state of neutron star matter could not anticipates the existence of neutron
stars in higher dimensions. Therefore, the modern equations of state of
neutron star matter derived from the microscopic calculations is a valid
equation of state in $4$\ dimensions for neutron stars, but for higher
dimensions, this equation of state should be modified. On the other hand, B.
C. Paul et al showed that for anisotropic stars, by considering higher
dimensions, the maximum mass of compact objects increases when dimensions of
spacetime increase in which however attains a maximum in $5-$dimensions,
thereafter it decreases as one increases number of extra dimensions (see
\cite{PaulCK2014}, for more details). As it was pointed out before, these
different results may be due to different equation of states or various
kinds of compact objects (anisotropic and isotropic). So it will be
interesting to study the equation of state of neutron star matter in higher
dimensions.

\section{Theory and observations \label{theory}}

Here, we compare our results of gravity's rainbow with observational data.
For this purpose, we present these results in table \ref{tab33}.

\begin{table*}[tbp]
\caption{Mass and radius of neutron star through the observations and
theory. }
\label{tab33}
\begin{center}
\begin{tabular}{cc}
\begin{tabular}{ccccccccc}
\hline\hline
& $Observations$ &  &  &  &  & $Theory$ &  &  \\
$name$ & ${M}\ (M_{\odot})$ & $R\ (km)$ &  &  & $H(\varepsilon)$ & ${M_{max}}%
\ (M_{\odot})$ & $R\ (km)$ &  \\ \hline\hline
$PSR\ J0348+0432$ & $2.01$ & $13(\pm 2)$ &  &  & $1.195$ & $2.01$ & $%
10.06(\pm 0.01)$ &  \\ \hline
$PSR\ J1614-2230$ & $1.97$ & $13(\pm 2)$ &  &  & $1.172$ & $1.97$ & $%
9.88(\pm 0.01)$ &  \\ \hline
$4U\ 1608-52$ & $1.74$ & $9.3(\pm 1)$ &  &  & $1.035$ & $1.74$ & $8.71(\pm
0.01)$ &  \\ \hline
\end{tabular}
&
\end{tabular}%
\end{center}
\end{table*}

As one can see, the results extracted in this theory match with the results
obtained through the observations, with the exception that the radius
obtained through gravity's rainbow are smaller than the radius of compact
objects. This is due to the fact that we have considered static neutron
stars in our studies, so the radius of neutron stars is smaller than the
radius of observational compact objects.

\begin{table*}[tbp]
\caption{Predicted radius for neutron star through the theory.}
\label{tab44}
\begin{center}
\begin{tabular}{cc}
\begin{tabular}{ccccccccc}
\hline\hline
& $Observations$ &  &  &  &  & $Theory$ &  &  \\ \hline\hline
$name$ & $M\ (M_{\odot})$ & $R\ (km)$ &  &  & $H(\varepsilon)$ & ${M\
(M_{\odot}})$ & $R\ (km)$ &  \\ \hline\hline
$J1748-2021B$ & $2.70$ & $unknown$ &  &  & $1.605$ & $2.70$ & $13.52(\pm
0.01) $ &  \\ \hline
$4U\ 1700-377$ & $2.40$ & $unknown$ &  &  & $1.425$ & $2.40$ & $12.00(\pm
0.01)$ &  \\ \hline
$PSR\ J1903+327$ & $1.67$ & $unknown$ &  &  & $0.994$ & $1.67$ & $8.37(\pm
0.01)$ &  \\ \hline
$Cen\ X-3$ & $1.49$ & $unknown$ &  &  & $0.888$ & $1.49$ & $7.48(\pm 0.01)$
&  \\ \hline
$PSR\ B1913+16$ & $1.44$ & $unknown$ &  &  & $0.856$ & $1.44$ & $7.20(\pm
0.01)$ &  \\ \hline
$PSR\ J0737-3039$ & $1.35$ & $unknown$ &  &  & $0.802$ & $1.35$ & $6.75(\pm
0.01)$ &  \\ \hline
$PSR\ J0737-3039B$ & $1.24$ & $unknown$ &  &  & $0.738$ & $1.24$ & $%
6.21(\pm0.01)$ &  \\ \hline
$SMC\ X-1$ & $1.04$ & $unknown$ &  &  & $0.738$ & $1.24$ & $6.21(\pm0.01)$ &
\\ \hline
\end{tabular}
&
\end{tabular}%
\end{center}
\end{table*}
\begin{figure}[tbp]
$%
\begin{array}{cc}
\epsfxsize=7cm \epsffile{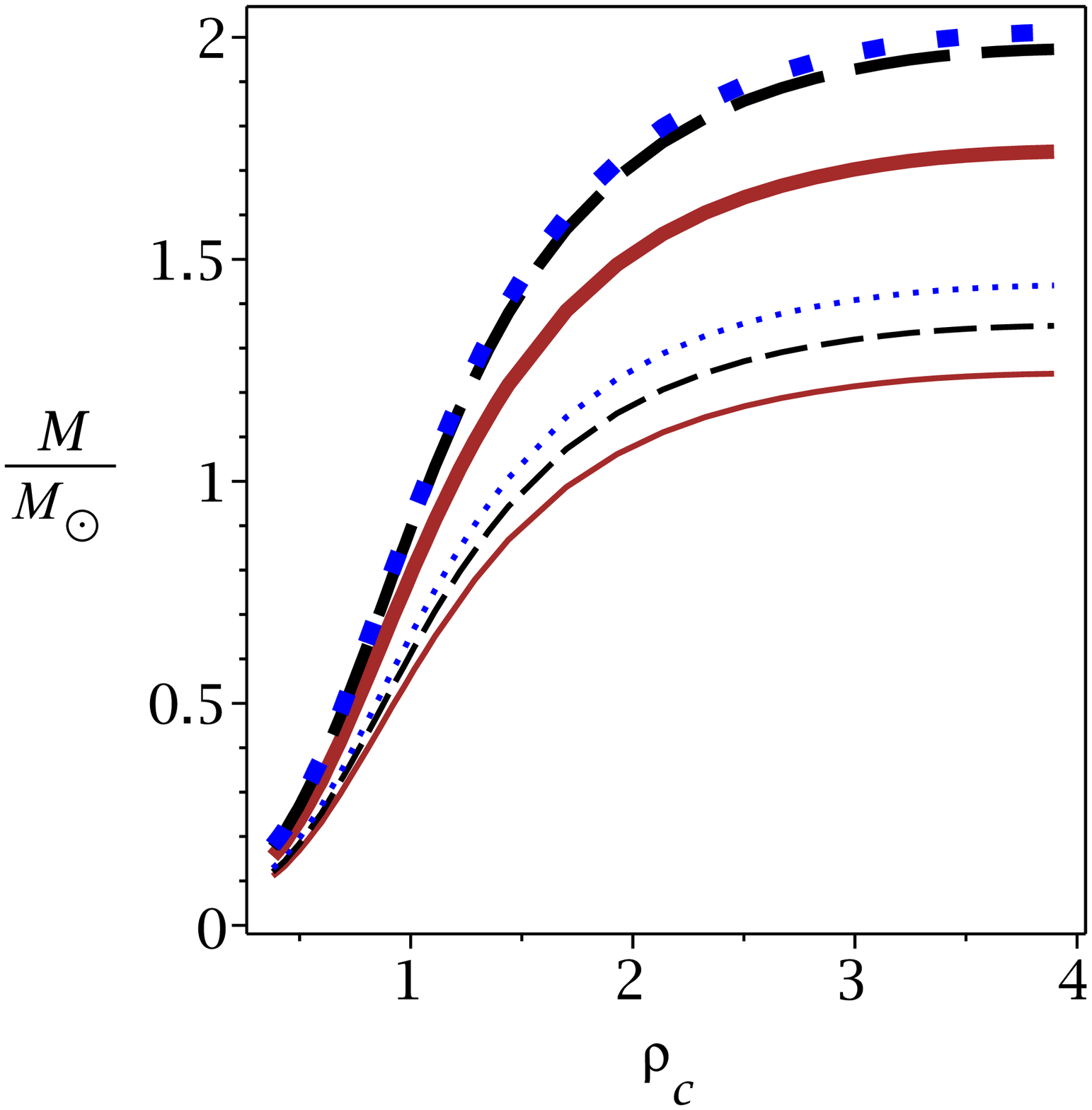} & \epsfxsize=7cm %
\epsffile{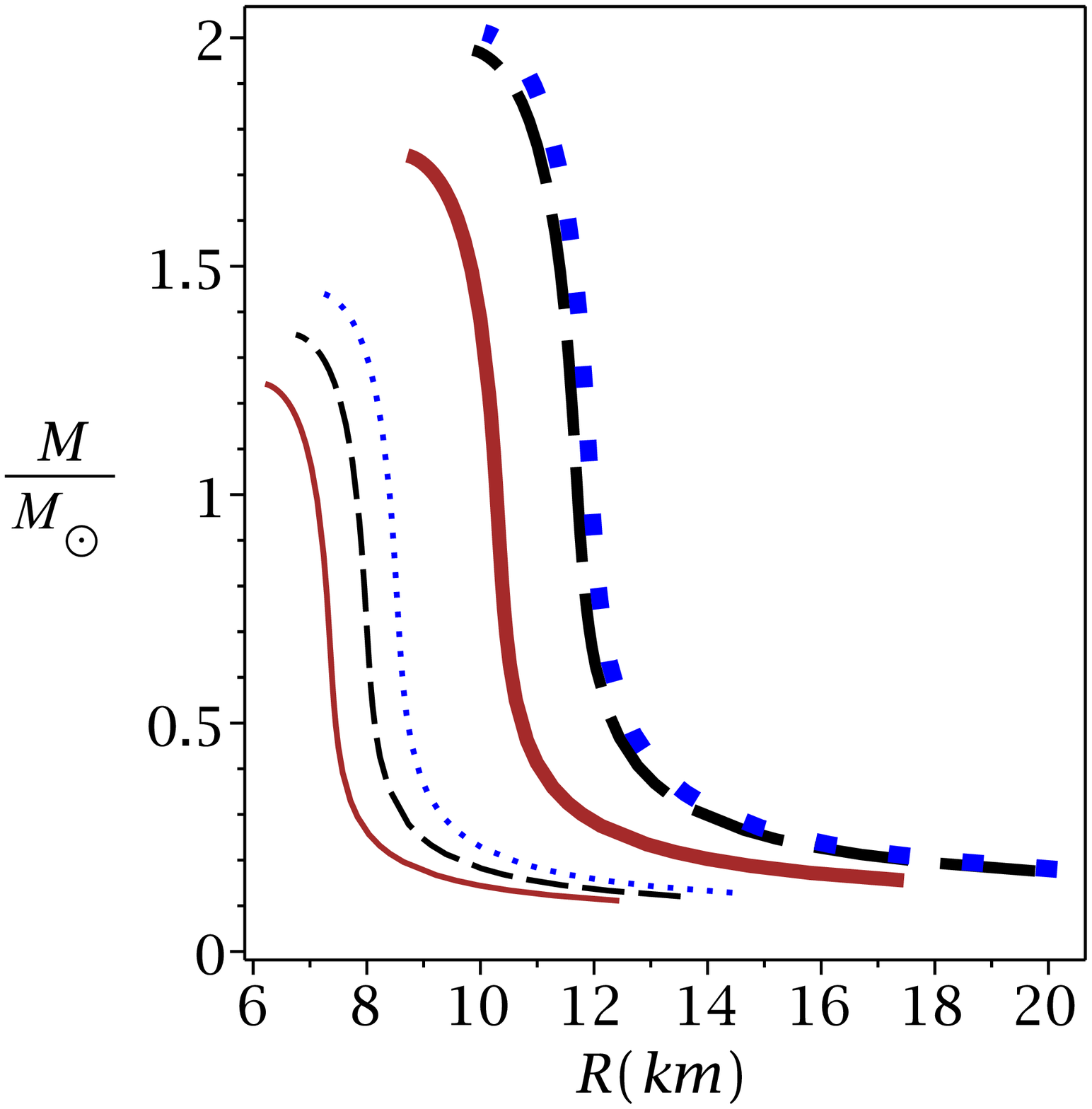}%
\end{array}
$%
\caption{Gravitational mass versus central mass density (left
diagram) and radius (right diagram), for $PSR\ J0737-3039B$
(continuous line), $PSR\ J0737-3039$ (dashed line), $PSR\
B1913+16$ (doted line), $4U\ 1608-52$ (bold-continuous line),
$PSR\ J1614-2230$ (bold-dashed line) and $PSR\ J0348-0432$
(bold-doted line).} \label{Fig13}
\end{figure}

Evidently for static cases of observational objects, our theory under
consideration predicts a set of radii which for cases such as J1748-2021B
and PSR J0348+0432, predicted radii are more than $13.52$ (km) and $10.06$
(km), respectively (see table \ref{tab44} and Fig. \ref{Fig13} for more
details).

\section{Conclusions \label{Conclusions}}

In this paper, by taking into account three models of gravity's rainbow with
$(3+1)$-dimensional spherically symmetric space-time, we studied the
hydrostatic equilibrium equation (HEE). Then, we generalized our solutions
to arbitrary $d$-dimensions ($d\geqslant 5$). Next, by employing obtained
HEE, we conducted a study regarding the structure of neutron stars.
Regarding microscopic calculations, we used the modern equations of state of
neutron star to obtain the properties of neutron star. We have studied the
effects of rainbow functions on the diagrams related to the mass-central
mass density relation and also the mass-radius relation of neutron star. We
showed that, for $H(\varepsilon )>1$, by increasing $H(\varepsilon )$, the
maximum mass of neutron star increases to more than $2M_{\odot }$ and for $%
H(\varepsilon )<1$, the maximum mass is a decreasing function of the $%
H(\varepsilon )$.

Then, we added the cosmological constant into this gravity and examined the
diagrams related to $M$-$\rho _{c}$ and $M$-$R$. It was pointed out that for
$\Lambda <10^{-14}\ m^{-2}$, the cosmological constant ($\Lambda $) does not
affect the maximum mass and radius of neutron stars. On the other hand, for $%
\Lambda >10^{-14}\ m^{-2}$, the mass versus radius of a neutron star may
present a quark-like star behavior (see Fig. \ref{Fig8}, for more details).

Next, we examined the effects of rainbow functions and the cosmological
constant on the other properties of neutron star such as; the Schwarzschild
radius, average density and strength of gravity. Also, we found a bound for
rainbow function, so that for the value less than $0.6$ ($H(\varepsilon
)<0.6 $), obtained average density for neutron stars was larger than of the
central density, therefore, valid values for rainbow function in case of
neutron stars are those that are larger than $0.6$ ($H(\varepsilon )>0.6$).
Another interesting result of this paper is the effects of the cosmological
constant on the so-called gravity strength. We found that for positive of
the cosmological constant, the strength of gravity is a decreasing function
of the cosmological constant which leads to decreasing the maximum mass of
neutron stars.

Also, we have obtained the gravitational redshift and showed that it was
modified in this gravity. The gravitational redshift of neutron stars in
this gravity was in the range $z\leq 0.566$ which is lower than the upper
bound on the surface redshift for a subluminal equation of state, i.e $%
z=0.851$ \cite{Haensel}. In addition, we have investigated the dynamical
stability and found that these stars were stable against the radial
adiabatic infinitesimal perturbations. Our results showed that, the neutron
stars with a mass more than $2M_{\odot }$, can exist in gravity's rainbow.
Then, we compared our results with the empirical data and saw that our
results were compatible with those of observations. Finally, using gravity's
rainbow, we obtained the radius of observational compact objects. We
concluded that it is possible to obtain the mass and radius of compact
objects by this theory through fine tuning.

One of the most important results of this paper is regarding determining and
finding an upper and a lower limitations for rainbow function. In other
words, gravity's rainbow was valid in the range of $0.6<H(\varepsilon )\leq
1.67$. Furthermore, this gravity predicts a maximum limitation for the mass
of neutron star which was about $2.81M_{\odot }$. Indeed, gravity's rainbow
covers all neutron stars and pulsars which have the maximum mass less than $%
2.81M_{\odot }$.

Using gravity's rainbow we obtained relation between the maximum mass of
neutron stars and the Planck mass. We showed that, this relation is modified
and it depends on rainbow function. It is notable that, in the absence of
rainbow function ($H(\varepsilon )=1$), this relation reduces to Einstein
gravity.

Considering the gravity's rainbow and the obtained modified TOV in this
gravity, we can investigate other compact objects such as white dwarf and
quark stars. In addition, it is worth studying the effects of other equation
of states on structure of the compact objects. On the other hand,
anisotropic compact objects \cite%
{HarkoM,BoehmerH,DonevaY,SunzuM,PaulD,NgubelangaM,MauryaG,RatanpalTP} in the
context of gravity's rainbow are also interesting subjects. We leave these
issues for future works.

\acknowledgments

The authors would like to thank anonymous referee for insightful comments.
We wish to thank Shiraz University Research Council. B. Eslam Panah
acknowledges K. Parvaneh and M. Momennia for helpful discussions. This work
has been supported financially by the Research Institute for Astronomy and
Astrophysics of Maragha, Iran.

\end{document}